\documentclass{aa}

\usepackage{graphicx}

\newlength{\figwidth}
\begin{document}

\title{Exploring the $P_{cyc}$ vs $P_{rot}$ relation with flux transport dynamo models of solar-like stars}
\titlerunning{Flux transport dynamo models of solar-like stars}

\author{L. Jouve \inst{1,2}, B.P. Brown\inst{3} \and A.S. Brun \inst{1,3}}
\authorrunning{Jouve, Brown, Brun}
\offprints{A.S. Brun\\ \email{sacha.brun@cea.fr}}

\institute{Laboratoire AIM, CEA/DSM-CNRS-Universit\'e Paris Diderot,
IRFU/SAp, 91191 Gif sur Yvette, France \and
D.A.M.T.P., Centre for Mathematical Sciences, University of Cambridge, Cambridge CB03 0WA, UK \and
JILA and Department of Astrophysical and Planetary Sciences, University of  Colorado, Boulder, CO 80309-0440, USA}

\date{Received ?? ; accepted ??}

\abstract{}{To understand stellar magnetism and to test the validity of the Babcock-Leighton flux transport mean field dynamo models with stellar activity observations}
{2-D mean field dynamo models at various rotation rates are computed with the STELEM code to
study the sensitivity of the activity cycle period and butterfly diagram to parameter changes and are compared 
to observational data. The novelty is that these 2-D mean field dynamo models incorporate scaling laws deduced 
from 3-D hydrodynamical simulations for the influence of rotation rate on the amplitude and profile of the meridional circulation. 
These models make also use of observational scaling laws for the variation of differential rotation with rotation rate.}
{We find that Babcock-Leighton flux transport dynamo models are able to reproduce the change in 
topology of the magnetic field (i.e. toward being more toroidal with increasing rotation rate) but 
seem to have difficulty reproducing the cycle period vs activity period correlation observed in solar-like stars if a monolithic 
single cell meridional flow is assumed. It may however be possible to recover the $P_{cyc}$ vs $P_{rot}$ relation with more complex meridional flows,
if the profile changes in a particular assumed manner with rotation rate.}
{ The Babcock-Leighton flux transport dynamo model based on single cell meridional circulation does not 
reproduce the $P_{cyc}$ vs $P_{rot}$ relation unless the amplitude of the meridional circulation is assumed to increase 
with rotation rate which seems to be in contradiction with recent results obtained with 3-D global simulations.}


\maketitle

\section{Introduction}
The Sun is a magnetic star that possesses a well defined cyclic activity period of 22 years.
It is believed that the origin of this magnetism and regular activity is linked to a global
scale dynamo operating in and at the base of the solar convection zone (Parker 1993). In order to model
and understand this global scale dynamo, a useful approach has been to make use of the mean field
dynamo theory (\cite{Mof78}; Krause \& Raedler 1980). This method has the advantage that it
only deals with the large scale magnetic field, assuming some parameterization of the underlying
small scale turbulence and magnetism. Computing those physical processes self-consistently would otherwise 
require costly full 3-D magnetohydrodynamical (MHD) simulations (Cattaneo 1999; Brun, Miesch \& Toomre 2004). 
Among the various mean field dynamo models, the Babcock-Leighton flux transport dynamo models have recently been the favored ones and have
demonstrated some success at reproducing solar observations assuming either an advection or a diffusion dominated
regime for the transport of field from the surface down to the tachocline (Dikpati \& Charbonneau 1999; Nandy \& Choudhuri 2002; Dikpati et al. 2004; 
Chatterjee, P. et al. 2004; Charbonneau 2005; Bonanno et al. 2005; Jouve \& Brun 2007; Yeates et al. 2008).

It is legitimate to ask if such models can be applied to other stars and lead to
the same agreement with the observations of for instance solar type stars that may possess
different rotation rates and activity levels. 

\subsection{Observational Constraints}

Observations of the magnetism of 'solar type' stars, i.e.
stars possessing a deep convective envelope and a radiative interior (late F, G, K and early M spectral type) are
becoming more and more available (Giampapa 2004) and may provide additional constraints on our understanding of 
global scale stellar dynamos. One difficulty of such observational programs is that they require long term observations
since stellar cycle periods are likely to be commensurate to the solar 11-yr sunspots cycle period (or 22-yr for a full cycle 
including two polarity reversals of the global poloidal field). 
Thanks to the data collected at Mount Wilson Observatory since the late 60's, such data is available (Wilson 1978; Baliunas et al. 1995).
Among the sample of 111 stars (including the Sun as a star) originally observed between the F2 to M2 spectral type, 
it is found that about 50\% of the stars possess a cyclic activity, with cycle (starspot) periods varying roughly between 5 to 25 yrs, 
i.e. between half to twice the sunspot cycle period. They further indicate that among the inactive stars of the
sample some are likely to be in a quiet phase (as was the Sun during the Maunder minimum). Overall activity cycles seems to
be more frequent for less massive stars K than for F stars. More recent observational programs have been pursued 
that now even provide information on the field topology as a function of the rotation rate, such as the one using 
the Espadons and Narval instruments and the Zeeman Doppler Imaging technique (Donati et al. 2006). Applying this observational
technique over a sample of four solar analogues with rotation rate varying from one to three times solar, Petit et al. (2008) have 
shown that the field amplitude increases as a function of the star's rotation rate and, more importantly, becomes more and more dominated by 
its toroidal component (modulo possible bias in the observational technique used). If such a trend is confirmed, 
i.e. that the field topology is becoming more toroidal with increasing rotation rate, it is a very important and instructive result and puts strong constraints on the dynamo models. 

The systematic analysis of stellar magnetism data revealed that for solar type stars there is a good correlation
between the cycle and rotation periods of the stars and that correlation is even stronger 
when using the Rossby number ($Ro = P_{rot}/\tau$) that takes into account the convection turnover time $\tau$ at the base
of the stellar convective envelope (Noyes et al. 1984; Soon et al. 1993; Hempelmann et al. 1995; Baliunas et al. 1996).
As the star rotates faster, its cycle period is found to be shorter.  Typically Noyes et al. (1984) found that $P_{cyc} \propto P_{rot}^n$, with 
$n = 1.25 \pm 0.5$. Based on an extended stellar sample Saar \& Brandenburg (1999) and Saar (2002) have argued that there is actually two branches when plotting the cycle period vs the rotation
period of the stars. They make the distinction between the primary (starspot) cycle and Gleissberg or grand minima 
type modulation of the stellar activity. For the active branch they found an exponent $n \sim 0.8$ and for the inactive stars $n\sim 1.15$ (Charbonneau \& Saar 2001). It is also found that this correlation breaks
at high rotation rate with the possible appearance of a super active branch. Further at very high rotation rate saturation of the X-ray luminosity seems to limit the validity of the scaling found 
at more moderate rotation rates (Pizzolato et al. 2003). For G type stars
this saturation is found for rotation rate above 35 ${\bf \rm km s^{-1}}$, for K type stars at about 10 ${\bf \rm km s^{-1}}$ and for M dwarfs around
3-4 ${\bf \rm km s^{-1}}$ (Browning 2008; Reiners, Basri \& Browning 2009). How stellar magnetic flux scales with rotation rate is thus also important
to understand since it is telling us how the magnetic field generated by dynamo action inside the stars emerges and imprints
the stellar surface (Rempel 2008) and if it actually saturates. 

\subsection{Possible theoretical interpretation}

Theoretical considerations to interpret stellar magnetism based on classical mean field $\alpha$-$\omega$ dynamo models  (Durney \& Latour 1976, Knobloch et al. 1981; Baliunas et al. 1996) 
naturally find correlation between rotation rate and stellar activity. In particular it is found that both magnetic field generation and the dynamo number $D$ 
(i.e. a Reynolds number characterizing the mean field $\alpha$ and $\omega$ dynamo effects used in the models) vary with the rotation period of the star. This is due to the fact that in these models both effects are sensitive to
the rotation rate of the star. The $\omega$-effect is a direct measure of the differential rotation $\Delta \Omega$ established
in the star. It is well known both theoretically and observationally that the differential rotation in the convective envelope of
solar-type stars is directly connected to the star's rotation rate $\Omega_0$ (Donahue et al. 1996; Barnes et al. 2005; Ballot et al. 2007;
Brown et al. 2008; K\"uker \& R\"udiger 2008). However, the exact scaling $n_r$ (i.e. $\Delta \Omega \propto \Omega_0^{n_r}$) is still a matter of 
debate among the observers and theoreticians, being sensitive to the observational techniques used and to the modelling approach.

The $\alpha$-effect used in mean field theory is related to helical turbulence and represents a measure or parameterization of the mean electromotive force (emf) (Moffatt 1978; Pouquet et al. 1976). It is thus also linked to the 
rotation rate of the star and the amount of kinetic helicity present in its convective envelope.
However in Babcock-Leighton flux transport dynamo models the standard $\alpha$-effect is replaced by a surface term linked to the
tilt of the active regions with respect to the east-west direction (i.e. Joy's law, Kosovichev \& Stenflo 2008). This tilt is thought
to be due to the action of the Coriolis force during the rise of the toroidal structures that emerge as active regions 
(D'Silva \& Choudhuri 1993). Recent 3-D simulations in spherical shells (Fan 2008; Jouve \& Brun 2007b, 2009) seem to indicate
that this is not the only effect responsible for the observed tilt and that the twist and arching of the toroidal structures as well as 
the continuous action of the surface convection during the emergence have some influence on the resulting tilt. Thus the
dependence with rotation rate of the Babcock-Leighton surface source term used in this class of mean field dynamo models is not straightforward
to assess and needs to be studied in detail. 

Another important ingredient in flux transport models is the large scale meridional circulation (MC) used to connect the surface source term generating
the poloidal field to the region of strong shear at the base of the convection zone (i.e. the tachocline) where it will be subsequently sheared
by the $\omega$-effect in order to close the global dynamo loop (i.e. $B_{pol} \rightarrow B_{tor} \rightarrow B_{pol}$) . The meridional flow (or "conveyor belt") 
thus plays an important role in setting the cycle period of the global dynamo in this class of flux transport model. As a direct consequence it is natural 
to ask how the meridional circulation amplitude and profile change with
the rotation rate and how these may influence the magnetic cycle period. 

Early work by Dikpati et al. (2001) have assumed that the amplitude of the 
meridional flow is proportional to the rotation rate of the star. With such scaling they seem to be able to reproduce some of the
rotation vs cycle period correlation observed in solar type stars. Charbonneau \& Saar (2001) have also studied the sensitivity of various
mean field dynamo models in reproducing stellar activity observations assuming for the Babcock-Leighton type that the meridional flow amplitude increases either as $\Omega_0$ or
as $\log(\Omega_0)$. Similarly Nandy (2004) and Nandy \& Martens (2007) have explored the solar-stellar connection with B-L mean field dynamo models and have
reached the same conclusion: only a positive scaling of the amplitude of meridional flows with the rotation rate can reconcile the models with observations of magnetism of solar-like stars.
However recent theoretical work by Ballot et al. (2007) and 
Brown et al. (2008) indicate that this is unlikely to be the right scaling. Instead 3-D simulations reveal that the amplitude of the meridional
flow weakens as the rotation rate is increased. 

In this paper we wish to revisit the solar-stellar connection in the light of
the latest theoretical and observational work done in the community. We have thus designed a series of Babcock-Leighton models
with the STELEM code (Emonet \& Charbonneau 1998, private communication; Jouve \& Brun 2007a) that incorporate recent 3-D MHD results. By doing so  
we will study how these mean field models, quite successful at modelling the solar cycle, perform to explain the observed magnetic activity, 
cycle and field topology of solar type stars at various rotation rates.
In \S 2 we discuss the 2-D mean field models used in this study and the way we incorporate 3-D results obtained with the ASH code (Clune et al. 1999; 
Brun et al. 2004). In \S 3 we present our series of stellar mean field dynamo models,  while in \S 4 we vary the input parameters to evaluate the robustness of our results. 
In \S5 we discuss and put in perspective our results.

\section{The Models and Method}

\subsection{The model equation}
To model the solar global dynamo operating in solar like stars, we use the hydromagnetic induction equation, governing the evolution of the 
magnetic field ${\bf B}$ in response to advection by a flow field ${\bf V}$ and resistive dissipation:

\begin{equation}
\frac{\partial {\bf B}}{\partial t}=\nabla\times ({\bf V} \times{\bf B})-\nabla\times(\eta\nabla\times{\bf B}) 
\end{equation}
where $\eta$ is the magnetic diffusivity.

 As we are working in the framework of mean-field theory, we express both magnetic and velocity fields as a sum 
of large-scale (that will correspond to mean field) and small-scale (associated with fluid turbulence) contributions. 
Averaging over some suitably chosen intermediate scale makes it possible to write two distinct induction equations for 
the mean and the fluctuating parts of the magnetic field. The mean-field equation reads:

\begin{eqnarray}
\frac{\partial {\bf \langle B\rangle}}{\partial t}&=&\nabla\times ({\bf \langle V\rangle} \times{\bf \langle B\rangle})
+\nabla\times\langle{\bf v'}\times {\bf b'}\rangle \nonumber \\
&-&\nabla\times(\eta\nabla\times{\bf \langle B\rangle})
\end{eqnarray}

where ${\bf \langle B\rangle}$ and ${\bf \langle V\rangle}$ refer to the mean parts of the magnetic and velocity fields and ${\bf v'}$ and ${\bf b'}$ 
to the fluctuating components. A closure relation is then used to express the mean electromotive force $\langle{\bf v'}\times {\bf b'}\rangle$ 
in terms of mean magnetic field, leading to a simplified mean-field equation. In this work we will replace the emf by a surface
Babcock-Leighton term (Babcock 1961; Leighton 1969; Wang et al. 1991; Dikpati \& Charbonneau 1999; Jouve \& Brun 2007a) as described in details below.

Working in spherical coordinates and under the assumption of axisymmetry, we write the total mean magnetic field {\bf B} 
and velocity field {\bf V} as (for the sake of simplicity we omit $\langle \, \rangle$ in the rest of the paper):
\begin{equation}
{{\bf B}}(r,\theta,t)=\nabla\times (A_{\phi}(r,\theta,t) \hat {\bf e}_{\phi})+B_{\phi}(r,\theta,t) \hat {\bf e}_{\phi}
\end{equation}
\begin{equation}
{{\bf V}}(r,\theta)={\bf v_{p}}(r,\theta) + r\sin\theta \, \Omega(r,\theta) \hat {\bf e}_{\phi},
\end{equation}

Note that our velocity field is time-independent since we will not assume any fluctuations in time of the differential 
rotation $ \Omega$ or of the meridional circulation ${\bf v_{p}}$.
Reintroducing this poloidal/toroidal decomposition of the field in the mean induction equation, we get two coupled 
partial differential equations, one involving the poloidal potential $A_{\phi}$ and the other concerning the toroidal field $B_{\phi}$. 

\begin{equation}
\label{eqA2}
\frac{\partial {A_{\phi}}}{\partial t}=\frac{\eta}{\eta_{t}} (\nabla^{2}-\frac{1}{\varpi^{2}})A_{\phi}-
R_{e}\frac{\bf{v}_{p}}{\varpi}\cdot\nabla(\varpi A_{\phi})+C_{s}S(r,\theta,B_{\phi})
\end{equation}

\begin{eqnarray}
\label{eqB2}
\frac{\partial {B_{\phi}}}{\partial t}&=&\frac{\eta}{\eta_{t}} (\nabla^{2}-\frac{1}{\varpi^{2}})B_{\phi}
+\frac{1}{\varpi}\frac{\partial(\varpi B_{\phi})}{\partial r}\frac{\partial (\eta/\eta_{t})}{\partial r} \nonumber  \\
&-&R_{e}\varpi {\bf v}_{p}\cdot\nabla(\frac{B_{\phi}}{\varpi})-R_{e}B_{\phi}\nabla\cdot{\bf v}_{p} \nonumber  \\ 
&+&C_{\Omega}\varpi(\nabla\times(\varpi A_{\phi}{\bf \hat{e}}_{\phi}))\cdot\nabla\Omega
\end{eqnarray}

where $\varpi=r\sin\theta$, $\eta_{t}$ is the turbulent magnetic diffusivity (diffusivity in the convective zone) and $S(r,\theta,B_{\phi})$ 
the Babcock-Leighton surface source term for poloidal field.
In order to write these equations in a dimensionless form, we choose as length scale the solar radius ($R_{\odot}$) and as time 
scale the diffusion time ($R_{\odot}^2/\eta_{t}$) based on the envelope diffusivity $\eta_{t} = 10^{11}\,\rm cm^2\rm s^{-1}$.
This leads to the appearance of three control (Reynolds numbers) parameters $C_{\Omega}=\Omega_{0}R_{\odot}^2/\eta_{t}$, $C_{s}=s_{0}R_{\odot}/\eta_{t}$ 
and $R_{e}=v_{0}R_{\odot}/\eta_{t}$ where $\Omega_{0}, s_{0}, v_{0}$ are respectively the rotation rate and the typical amplitude of  
the surface source term and of the meridional flow.

Equations $\ref{eqA2}$ and $\ref{eqB2}$ are solved with the STELEM code (Emonet \& Charbonneau 1998, private communication) in an annular meridional plane with the colatitude $\theta$ $\in [0,\pi]$ and the 
radius (in dimensionless units) $r \in [0.6,1]$ i.e. from slightly below the tachocline ($r=0.7$) up to the solar surface.
The STELEM code has been thoroughly tested and validated thanks to an international mean field dynamo benchmark involving 8 different codes (Jouve et al. 2008). 
At $\theta=0$ and $\theta=\pi$ boundaries, both $A_{\phi}$ and $B_{\phi}$ are set to 0. Both $A_{\phi}$ and $B_{\phi}$ are set 
to $0$ at $r=0.6$. At the upper boundary,  we smoothly match our solution to an external potential field, i.e. we have vacuum 
for $r \geq 1$.
As initial conditions we are setting a confined dipolar field configuration, i.e. the poloidal field is set to $\sin\theta / r^{2}$ in 
the convective zone and to $0$ below the tachocline whereas the toroidal field is set to $0$ everywhere.

\subsection{The physical ingredients}

\begin{figure}[!h]
  \centering
\includegraphics[width=8cm]{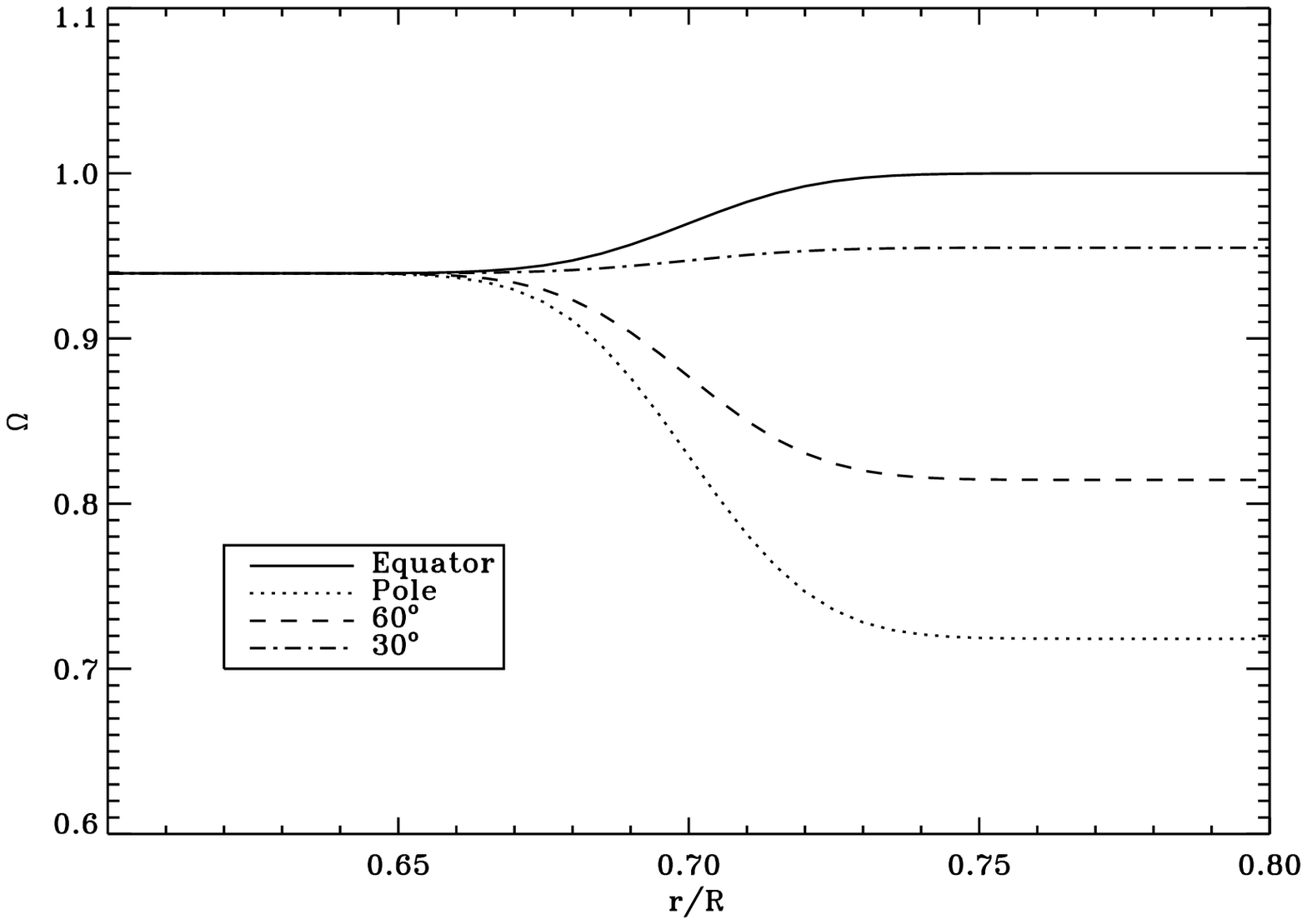}
\includegraphics[width=8cm]{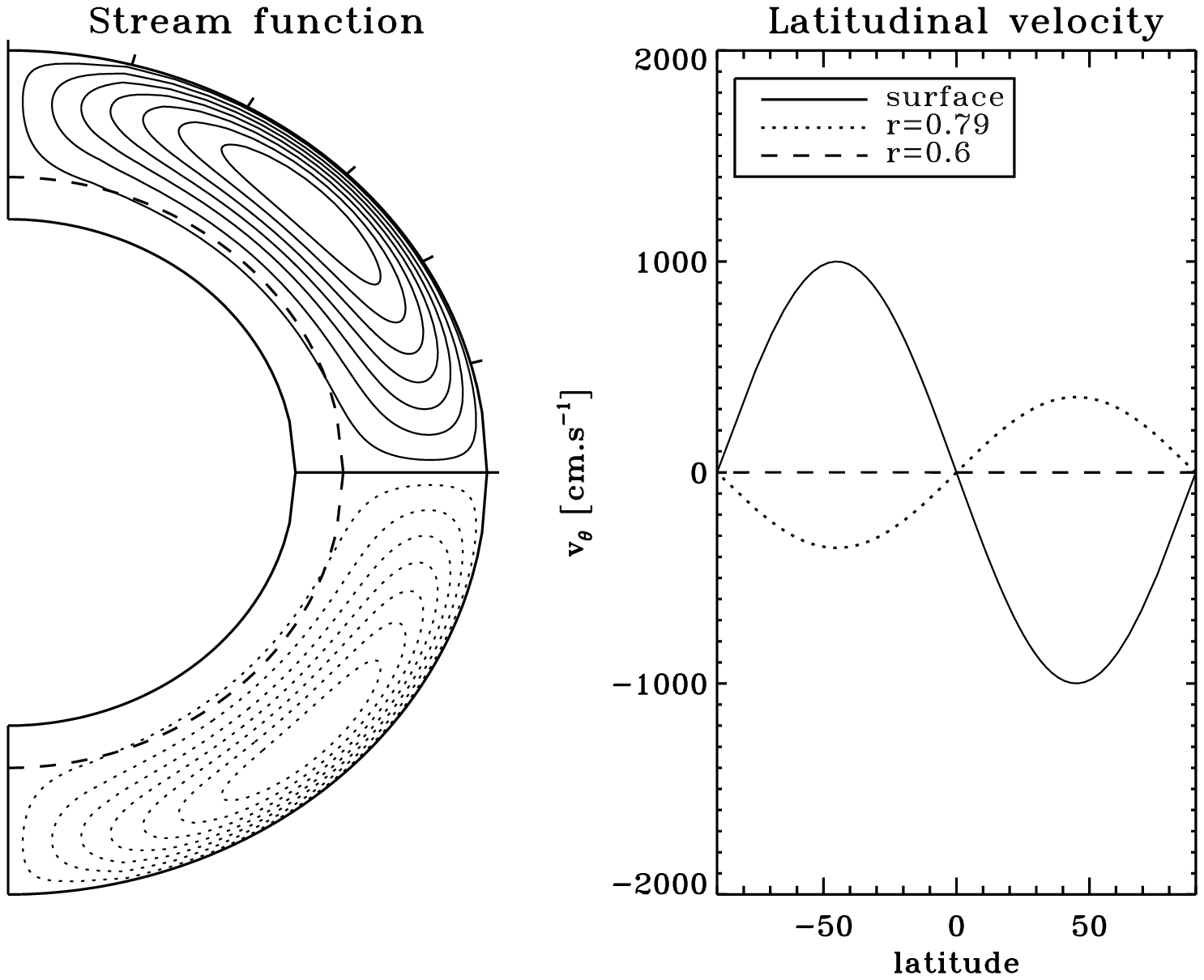}
\caption{Top panel: differential rotation profile used in the models G0.25 to G10 discussed section 3. 
We zoomed on the tachocline region between $r=0.6$ and $r=0.8$. Bottom panel: meridional circulation used in the models G0.25 to G10 discussed section 3. Plain (dashed) lines indicate counterclockwise (clockwise) circulation.} 
\label{domega}
\end{figure}

The rotation profile used in the series of models discussed in this work captures some realistic aspects of the Sun's angular velocity, deduced from helioseismic inversions (Thompson et al. 2003). We are thus 
assuming a solid rotation below $0.66$ and a differential rotation above the interface (cf. top panel of Figure \ref{domega}). 

\begin{eqnarray}
{\Omega(r,\theta)}&=&\Omega_{c}+\frac{1}{2}[1+erf(2\frac{r-r_{c}}{d_{1}})] \nonumber \\
& &(\Omega_{Eq}+a_{2}\cos^2\theta+a_{4}\cos^4\theta-\Omega_{c})
\end{eqnarray}

with $\Omega_{Eq}=1$,  $\Omega_{c}= 0.93944$,  $r_{c}=0.7$, $d_{1}=0.05$, $a_{2}=-0.136076$ and $a_{4}=-0.145713$.
With this profile, the radial shear is maximal at the tachocline. As already discussed in the introduction both observations (Donahue et al. 1995; Barnes et al. 2005) 
and 3-D numerical simulations (Ballot et al. 2007; Brown et al. 2008) indicate that the inner rotation profile is
likely to vary with the rotation rate. In the models discussed in section 3 we will keep the profile of differential rotation fixed, varying only the rotation rate $\Omega_0$ of the models.
We will study the influence of varying the differential rotation profile in section 4.1.

We assume that the diffusivity in the envelope $\eta$ is dominated by its turbulent contribution whereas in the stable interior 
$\eta_{\rm c} \ll \eta_{\rm t}$. We smoothly match the two different constant values with an error function which enables us to 
quickly and continuously transit from $\eta_{\rm c}$ to $\eta_{\rm t}$ i.e.

\begin{equation}
\eta(r)=\eta_{\rm c}+\frac{1}{2}\left(\eta_{\rm t}-\eta_{\rm c}\right)\left[1+{\rm erf}\left(\frac{r-r_{\rm c}}{d}\right)\right],
\label{eqeta}
\end{equation}

\noindent with  ${\eta_{\rm c}}=10^9 \,\rm cm^2\rm s^{-1}$ and $d=0.03$.

In Babcock-Leighton (BL) flux-transport dynamo models, the poloidal field owes its origin to the tilt of magnetic loops emerging at 
the solar surface. Thus, the source has to be confined to a thin layer just below the surface and since the process is fundamentally 
non-local, the source term depends on the variation of $B_{\phi}$ at the base of the convection zone (CZ). The expression is

\begin{eqnarray}
S(r,\theta,B_{\phi})&=& \frac{1}{2}[1+erf(\frac{r-r_{2}}{d_{2}})][ 1-erf(\frac{r-1}{d_{2}})] \nonumber \\ 
&\times&[1+({\frac{B_{\phi}(r_{c},\theta,t)}{B_{0}}})^{2}]^{-1}\cos\theta \sin\theta B_{\phi}(r_{c},\theta,t) \nonumber
\end{eqnarray}

\noindent where $r_{2}=0.95$, $d_{2}=0.01$, $B_{0}=10^5 \, \rm G$.

In these BL flux-transport dynamo models, meridional 
circulation is used to link the two sources of the magnetic 
field namely the base of the CZ and the solar surface.  The 
structure of the meridional circulations in the Sun is less 
constrained than the internal differential rotation, as they 
possess slower flows that are difficult to detect 
helioseismically.  The nature of these flows in other stars is 
even less constrained.  It is generally believed that in the Sun, 
the meridional circulations are likely to be composed largely of a 
single cell in each hemisphere, and recent solar simulations 
(Miesch et al. 2008) find such single-celled structures when the 
 global-scale flows are averaged over long intervals but multi-cellular profiles can also be realized.  
In more rapidly rotating suns, 3-D models presently indicate that the 
 circulations are likely to be multi-celled in both radius and 
 latitude (Ballot et al. 2007; Brown et al. 2008). 
In the series of models discussed in section 3 of this paper, 
we restrict ourselves to Babcock-Leighton flux-transport models 
that have a large single cell per hemisphere. We choose this
simple profile because it is widely used in the community, 
although some studies have considered more complex profiles (Bonanno et al. 2005; Jouve \& Brun 2007).
We are deferring to section 4.2 the analysis of complex multi-cellular profiles.
As in the Sun, the meridional circulations are 
directed poleward at the surface and here they vanish at the 
bottom boundary ($r=0.6$) and thus penetrate a little beneath the 
tachocline into the radiative interior (cf. bottom panel of Figure \ref{domega}). To model the single cell meridional circulation
we consider a stream function with the following expression:

\begin{equation}
\psi(r,\theta)=-\frac{2}{\pi}\frac{(r-r_{b})^2}{(1-r_{b})}\sin\left(\pi\frac{r-r_{b}}{1-r_{b}}\right)\cos\theta\sin\theta,
\label{eqpsi}
\end{equation}

\noindent which gives, through the relation  ${\bf v_{\rm p}}=\nabla \times(\psi \hat {\bf e}_{\phi})$, the following components of the meridional flow

\begin{eqnarray}
v_{r}&=&-\frac{2(1-r_{\rm b})}{\pi r}\frac{(r-r_{\rm b})^2}{(1-r_{\rm b})^2} \nonumber \\
& \times&\sin\left(\pi\frac{r-r_{\rm b}}{1-r_{\rm b}}\right)(3\cos^2\theta-1), 
\end{eqnarray}
\begin{eqnarray}
v_{\theta}&=&[\frac{3r-r_{\rm b}}{1-r_{\rm b}} \sin\left(\pi\frac{r-r_{\rm b}}{1-r_{\rm b}}\right) \nonumber \\
&+&\frac{r\pi}{1-r_{\rm b}}\frac{(r-r_{\rm b})}{(1-r_{\rm b})} \cos\left(\pi\frac{r-r_{\rm b}}{1-r_{\rm b}}\right)] \nonumber \\
	  & \times &\frac{2(1-r_{\rm b})}{\pi r}\frac{(r-r_{\rm b})}{(1-r_{\rm b})}\cos\theta\sin\theta, 
\end{eqnarray}
\noindent with $r_{\rm b}=0.6$. 

\section{Flux transport models as a function of rotation rate}

\subsection{Previous results}

Jouve \& Brun (2007a) have done a systematic study of the influence on the cycle period of the various parameters of Babcock-Leighton flux transport models. 
For the reference model possessing only one cell per hemisphere they confirm the results of Dikpati \& Charbonneau (1999) that the cycle period is strongly dependent 
on the meridional circulation amplitude. A least square fit through the ensemble of models they computed leads to the following dependency:

\begin{equation}\label{pcyc_theo}
P_{cyc} = v_0^{-0.91} s_0^{-0.013} \eta^{-0.075} \Omega_0^{-0.014} 
\end{equation}

We can thus expect that as we increase the rotation rate the cycle period will shorten if all the other parameters are kept constant but unfortunately not as fast as what is observed (Noyes et al. 1984; Saar 2002). However 
we may also expect the meridional circulation to depend on the rotation rate, and the sign of the variation of $v_0$ with $\Omega_0$ will most certainly dominate
over the weak dependence of $P_{cyc}$ on $\Omega_0$. It is thus crucial to assess how the meridional circulation amplitude and profile vary with the rotation rate $\Omega_0$.
We will consider in the following sections both various amplitude for the meridional circulation at a given rotation rate and more complex profiles with several cells per hemisphere as was done in Jouve \& Brun (2007a),
to quantify how this may affect the stellar activity cycle.

\subsection{Implementing results from 3D simulations of rapidly rotating stars in our mean field dynamo models}

In order to get some physical insight on how the meridional circulation varies with the rotation rate we rely on recent 3-D numerical simulations of solar convection performed
with the ASH code (Brun \& Toomre 2002, Ballot et al. 2007, Brown et al. 2008). Indeed there are no reliable observations of the meridional circulation in stars except for the Sun near the surface. 
The theoretical work of Brown et al. (2008), that studies the influence of the rotation rate on solar like stars (i.e. with the same thickness for the convection zone as our current Sun), will provide the required scaling. 
In this work several 3-D purely hydrodynamical models have been computed at various rotation rates (from 1 to 10 times solar) and the redistribution of angular
momentum, heat and energy assessed systematically to explain the differential rotation and meridional circulations achieved in the simulations. 
Figures 10 and 12 of Brown et al. (2008) are particularly relevant to our study and we have used the data from all the models reported in these figures
plus more slowly rotating cases (i.e. with a rotation rate lower than the solar rate) that are currently being studied (Brown 2009).

\begin{table}[!h]
\begin{center}
\begin{tabular}{|c|c|c|c|c|c|} \hline 
\bf Case    & \bf $\Omega_{0}/\Omega_{S}$ & \bf MCKE/$10^4$ & $\bf v_\theta^{scale}$ & ${\bf R_e}/10^2$ & ${\bf C_\Omega}/10^5$ \\ \hline
G0.25   &  0.25   &  23.4   &  28.7      &       20.1    &    0.351  \\ \hline
G0.5   &  0.5    &  16.2    &  23.9      &       16.7    &    0.702  \\ \hline
G0.6   &  0.6   &  12.3     &  20.9      &       14.6    &    0.842  \\ \hline
G0.75   &  0.75    &  4.75    &  12.9      &       9.06     &    1.05  \\ \hline
G0.9   &  0.9      &  3.09   &  10.4      &       7.29     &    1.26  \\ \hline
G1     &  1   &  2.84       &  10.0      &       7.00     &    1.40  \\ \hline
G1.25    &  1.25    &  2.18    &  8.76       &       6.13     &    1.75  \\ \hline
G1.5    &  1.5   &  1.98      &  8.36    &       5.85     &    2.11  \\ \hline
G1.75    &  1.75      &  1.68  &  7.70       &       5.39     &    2.46  \\ \hline
G2       &  2      &  1.52     &  7.33      &       5.13     &    2.81  \\ \hline
G3       &  3      &  1.11     &  6.25       &       4.38     &    4.21  \\ \hline
G4       &  4      &  0.767      &  5.20       &       3.64     &    5.62  \\ \hline
G5       &  5      &  0.711      &  5.01       &       3.51     &    7.02  \\ \hline
G6       &  6      &  0.626      &  4.70       &       3.29     &    8.42  \\ \hline
G7       &  7     &  0.497       &  4.18       &       2.93     &    9.83  \\ \hline
G10      &  10    &  0.359       &  3.56       &       2.49     &    14.0  \\ \hline
\end{tabular}  
\caption{Summary of the different cases studied and the values of $R_e$ and $C_\Omega$ used in the 2D Babcock-Leighton models.
The value of $\eta_t$ is fixed to $10^{11} \,\, \rm cm^2.s^{-1}$ and the value of $s_0$ is fixed to $50 \,\, \rm cm.s^{-1}$.
The stellar rotation rate $\Omega_{0}$ is given in terms of the solar rotation rate $\Omega_{S}$.
$v_\theta^{scale}$ represents the amplitude of $v_\theta$ at the surface at $45^{o}$ and is imposed equal to $10 \,\, \rm m.s^{-1}$ for case G1,
representing the reference solar mean field dynamo model.}
\label{table_A}
\end{center}
\end{table}

\begin{figure}[!h]
  \centering
\includegraphics[width=7.cm]{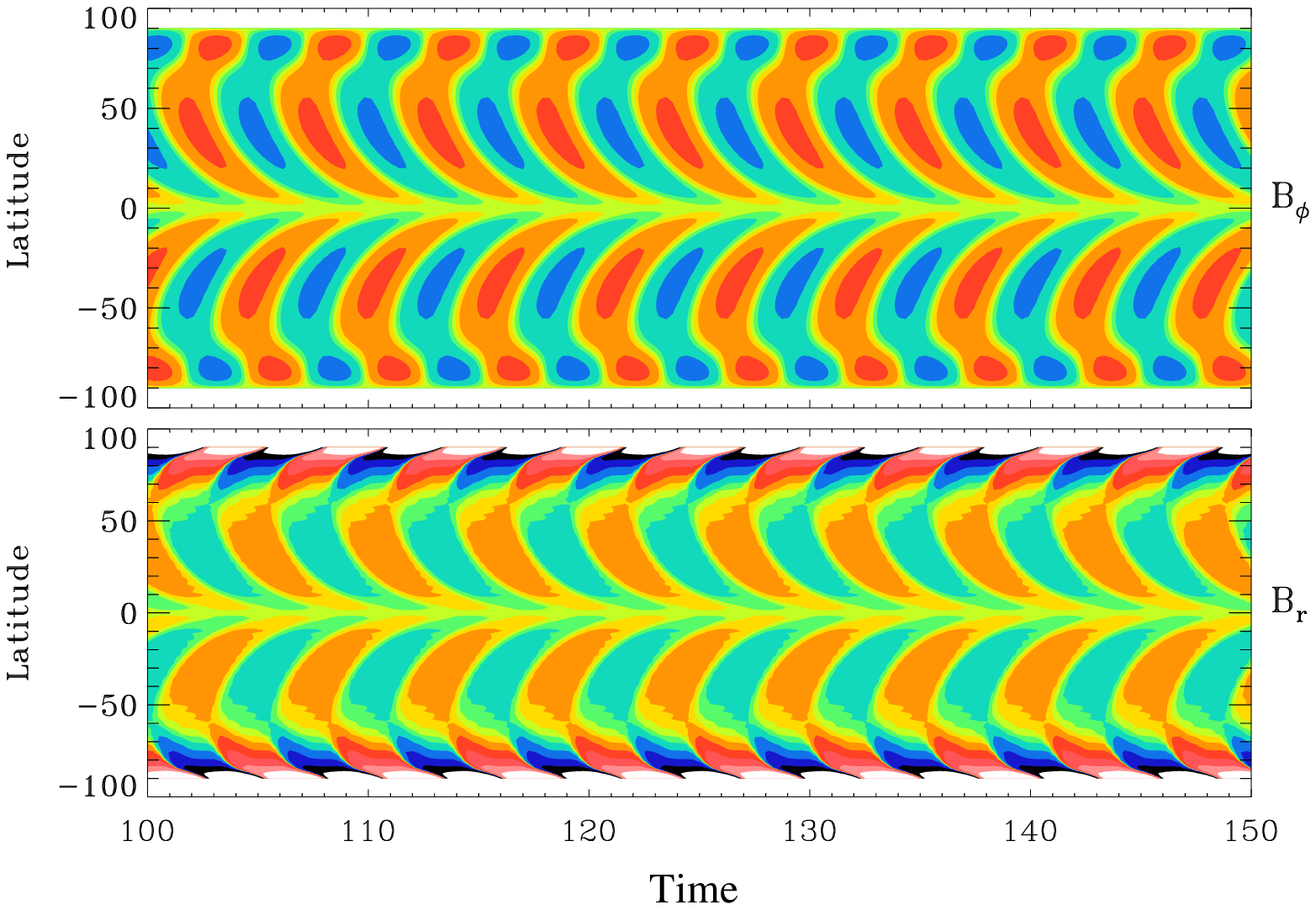}
\includegraphics[width=7.cm]{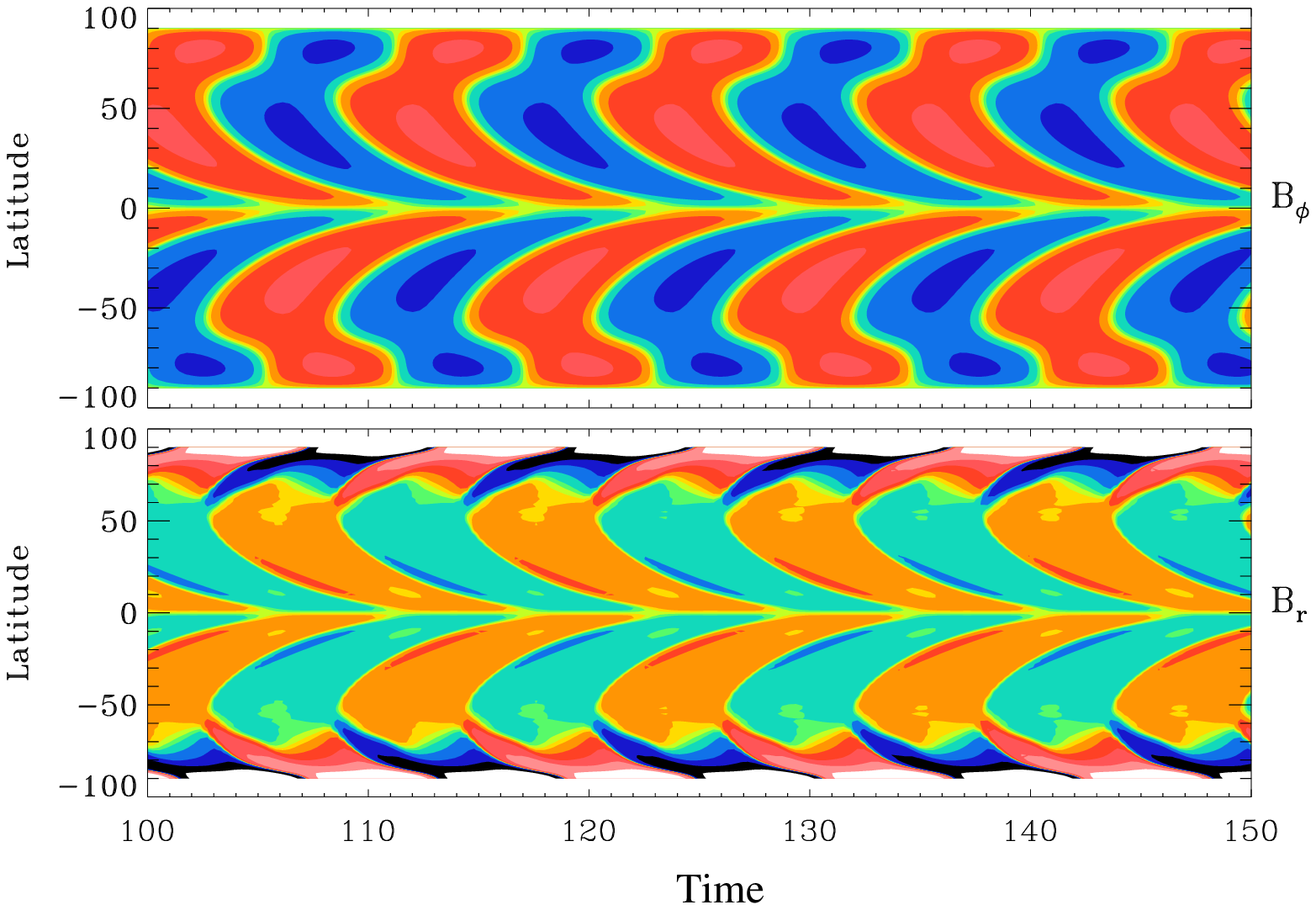}
\includegraphics[width=7.cm]{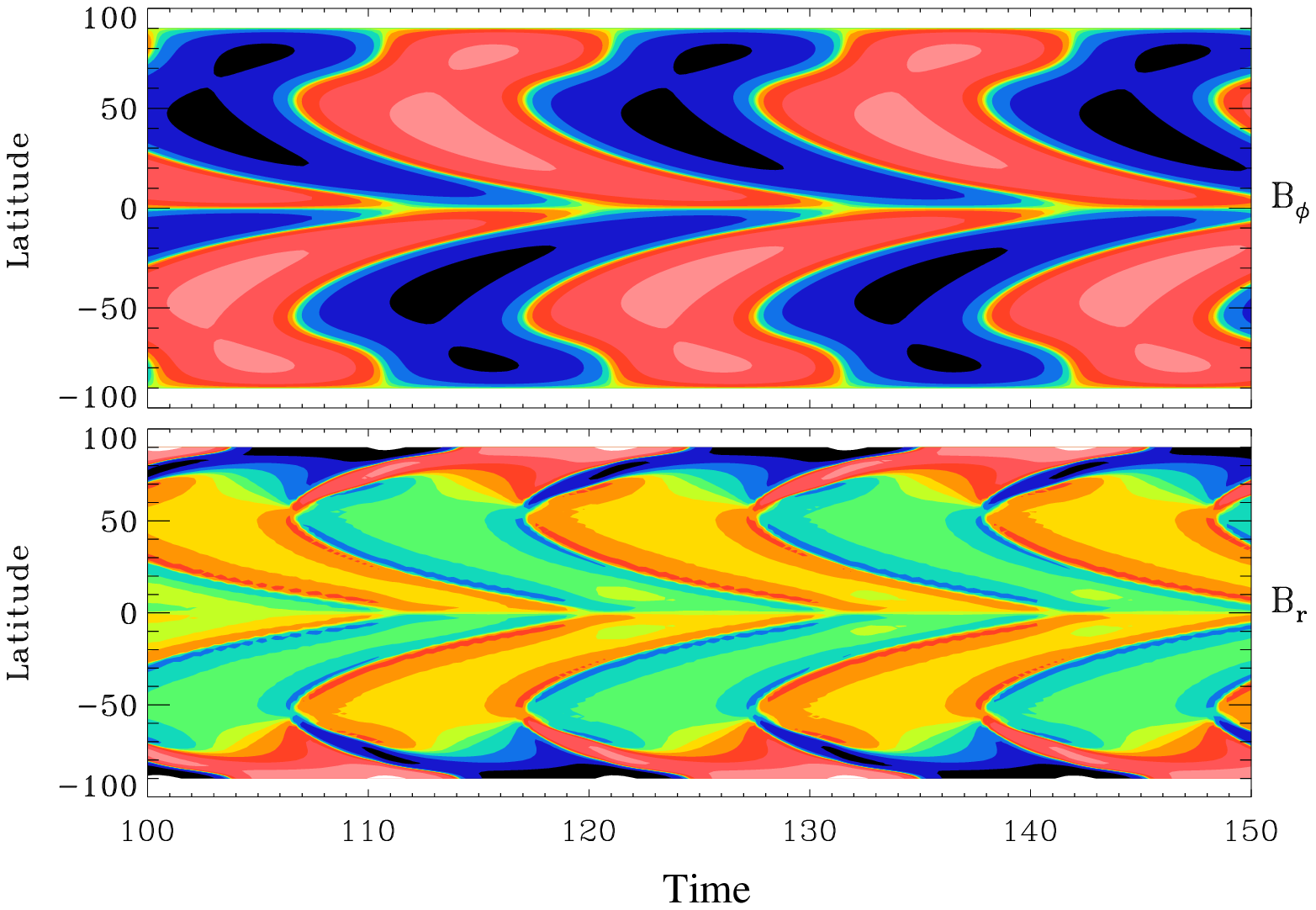}
\includegraphics[width=7.cm]{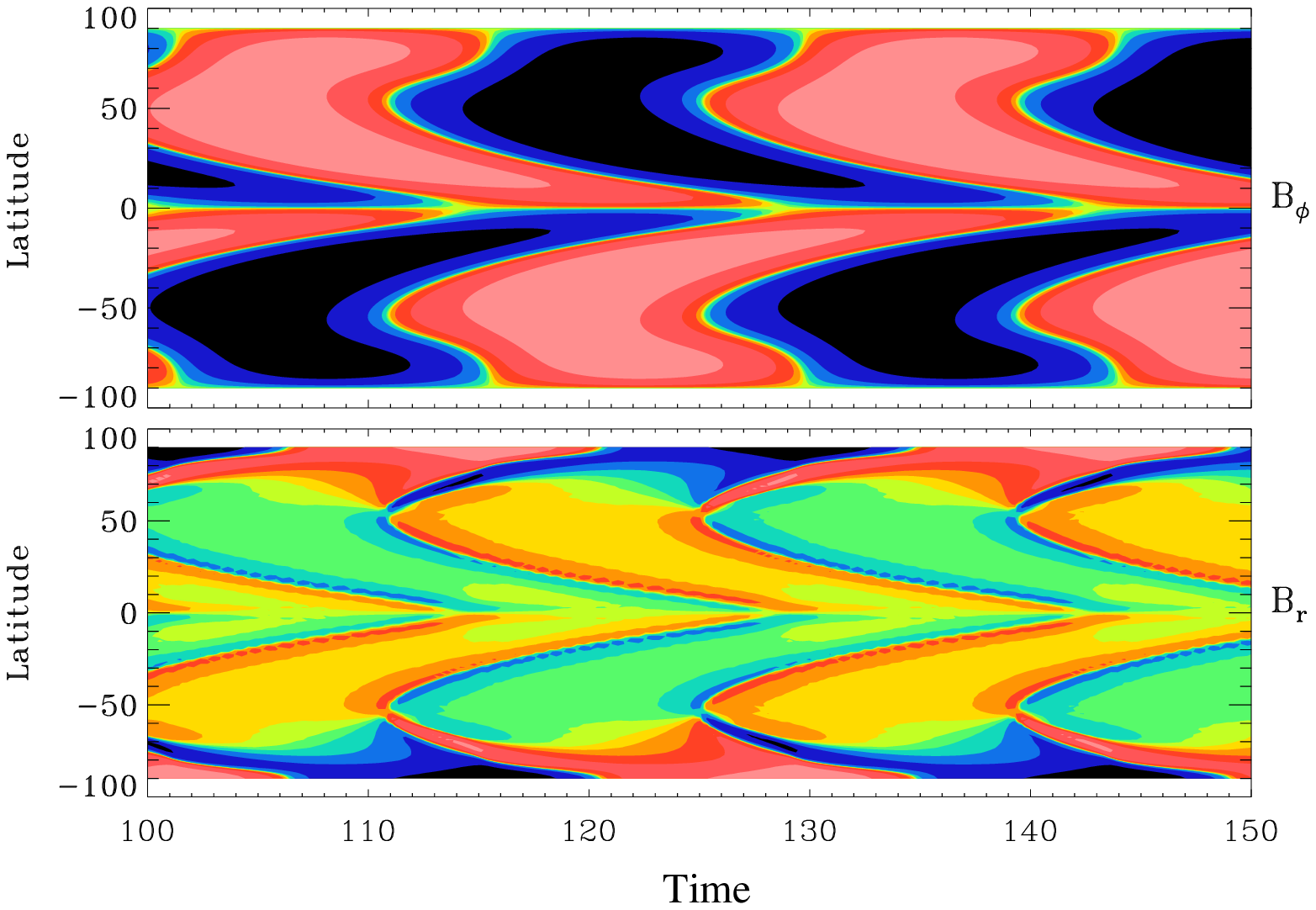}
\caption{Butterfly diagrams for cases G0.5, G1, G5 and G10. The top panel shows the time-latitude cut of the toroidal field at the base of the 
convection zone and the bottom one represents the evolution of the radial field at the surface. Time is in years. The scale of the color 
table is the same for all: between $-5\times10^{3}\,\, \rm G$ and $5\times10^{3}\,\, \rm G$ for the radial field and between 
$-5\times10^{5} \,\, \rm G$ and $5\times10^{5}\,\, \rm G$ for the toroidal field.} 
\label{fig_butterfly}
\end{figure}

We have thus scaled the meridional circulation by taking into account the way meridional circulation kinetic energy (MCKE)
 is found to change with rotation in the published 3-D models of Brown et al. (2008). 
We are confident that this scaling for MCKE still holds in the magnetized version since recent 
3-D models of dynamo action in rapidly rotating suns find that the weak meridional circulations are essentially unmodified by the strong magnetic fields that are generated (Brown et al. 2009). 
In contrast, the differential rotation appears to be substantially different 
from that found in hydrodynamic cases, and its scaling with 
rotation rate is sensitive to magnetic fields that are realized. We will thus not use the scaling
of $\Delta \Omega$ with $\Omega_0$ found in Brown et al. (2008) when we will vary the differential
rotation profile in our models but instead rely on observations (see section 4.1 below). 
        
In Table \ref{table_A}, we list the solar-like mean field dynamo models that we have computed at various rotation rate along with  
the associated Reynolds numbers $R_e$ and $C_\Omega$ that we have used in our 2-D models.  
A quick look at the table already reveals that the meridional circulation amplitude decreases with rotation rate. This may come as a surprise as one could have expected
faster meridional flows as $\Omega_0$ is increased. However a careful study, via the longitudinal component of the vorticity equation, of the maintenance of the meridional circulation (which arises from small unbalance between large forces; 
see for instance Miesch 2005; Brun \& Rempel 2008), reveals that it actually weakens with faster rotation rate.
It is thus likely given the strong negative dependency of $P_{cyc}$ with $v_0$ that
as we increase the rotation rate, the cycle period will increase rather than decrease as observed.  However it is not obvious to anticipate the resulting field topology of the various models without actually computing them.

Incorporating the values of the dynamo numbers $R_e$ and $C_\Omega$ for the different stars in our flux-transport BL dynamo models enables us to 
study the behaviour of the mean magnetic field in such stars in the framework of mean-field theory. All the dynamo models we have studied give cyclic behaviours with 
different periods and different intensity of the mean toroidal and poloidal magnetic fields.
We wish to stress that even though the naming scheme of the models discussed in this paper is similar to that used in Brown et al. (2008), the models are distinct since the ones discussed in this work are 2-D not 3-D.

\begin{figure}[!h]
  \centering
\includegraphics[width=0.95\linewidth]{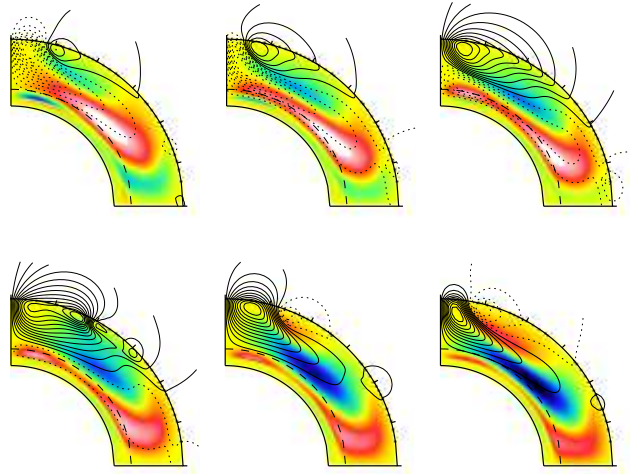}
\caption{Representative solution of a BL model applied to a 
      rapidly rotating star. This is case G5, with a rotation rate $\Omega_0$ of 
      5 $\Omega_{S}$ . We show the evolution of the toroidal field (filled colour 
      contours, with polarity indicated: red, positive; blue, negative) 
      and of the poloidal field (contour lines) over a period of 20.8  
      years, or half a magnetic cycle in this simulation.  A potential 
      field solution to the poloidal field is extrapolated from the 
      surface to 1.2$R_\odot$.} 
\label{fig_solution}
\end{figure}

Figure \ref{fig_butterfly} gives examples of 4 different butterfly diagrams obtained for more or less rapidly rotating stars. This figure shows the 
evolution in time of the toroidal magnetic field at the bottom of the CZ and of the radial field at the top of the domain. 
We note that the behaviour of the field is very similar in all these models, except that the period is clearly varying as well as the 
intensity of the magnetic field (the scale of the color table is the same for all panels). Indeed, we note that the magnetic cycle period is 
increased when the rotation rate is increased. This is a direct consequence of the fact that the cycle period is almost inversely proportional 
to the amplitude of the meridional flow in BL models. 

In order to illustrate the temporal evolution of the magnetic field in the meridional plane, we display on Figure \ref{fig_solution} a temporal sequence
over one starspot cycle for model G5. The advection of the field by the single-celled meridional circulation is evident confirming that our dynamo model is
dominated by advection and not by diffusion. We wish to stress here that our particular choice of advection-dominated regime is
not the only possibility at hand to model the transport occurring in the solar convection zone and the manner that the models link 
the Babcock-Leighton surface source term to the deep tachocline. For instance Chatterjee et al. (2004), K\"{a}pyl\"{a} et al. (2006), 
Yeates et al. (2008), Guerrero \& Gouveia Dal Pino (2008) have studied how transporting the poloidal field from the surface to the tachocline 
via turbulent diffusion or turbulent pumping influence the butterfly diagram, the polarity of the field and the cycle period. We intend to look at
these solutions in the near future. 

Returning to Figure \ref{fig_solution} we also note that the poloidal field tends to be more
concentrated near the polar region in agreement with the stronger polar branch seen in the corresponding butterfly diagram.
Indeed in Figure \ref{fig_butterfly} we clearly note that the magnetic intensity is increased when the rotation rate is increased, 
which is in agreement with the observations. 
We finally also note on the poloidal field that the contrast between the polar intensity and the 
intensity close to the equator is diminished when the rotation rate is increased.

\subsection{Field topology and poloidal vs toroidal field ratio}

\begin{figure}[!h]
  \centering
\includegraphics[width=6.8cm]{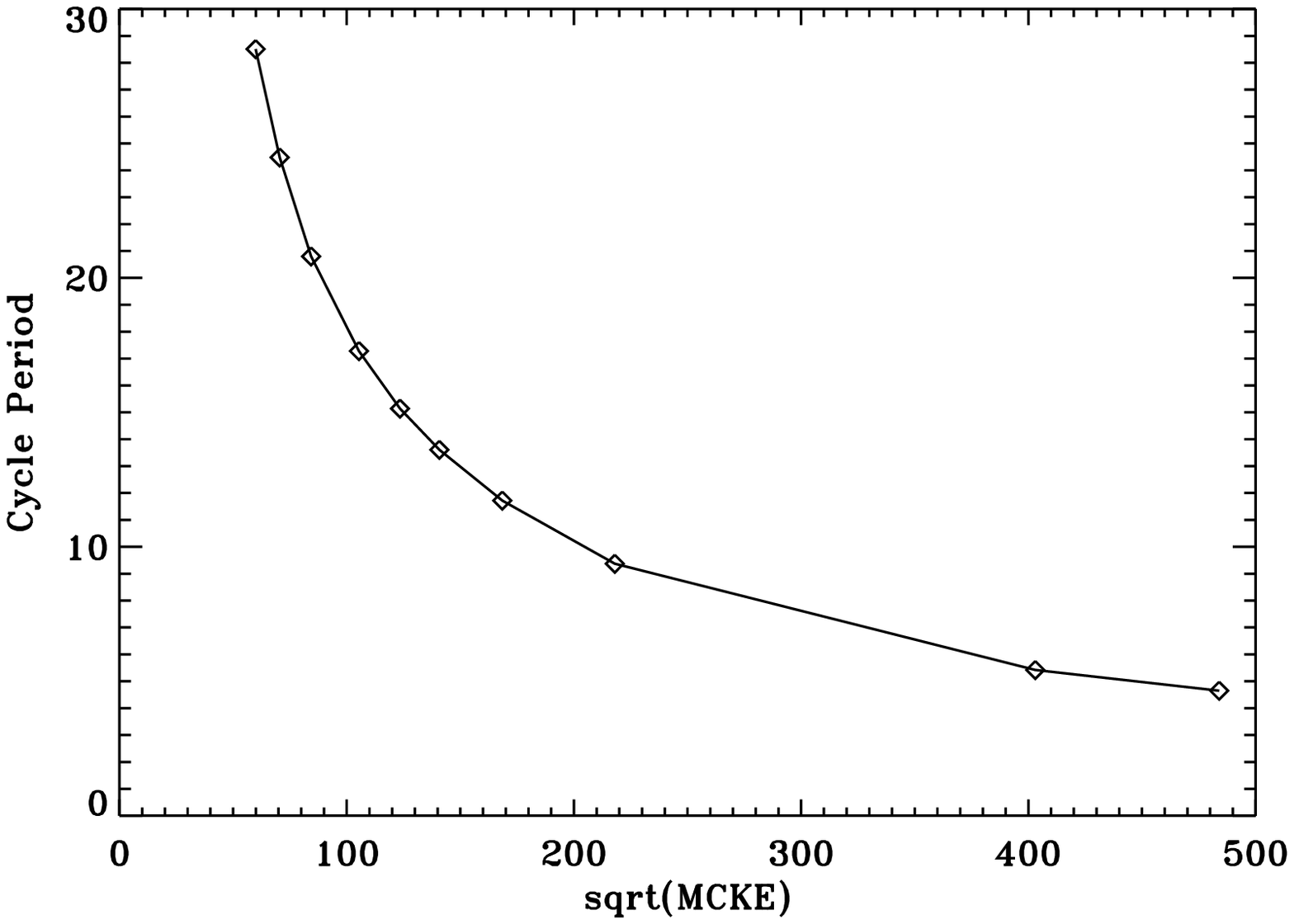}
\includegraphics[width=6.8cm]{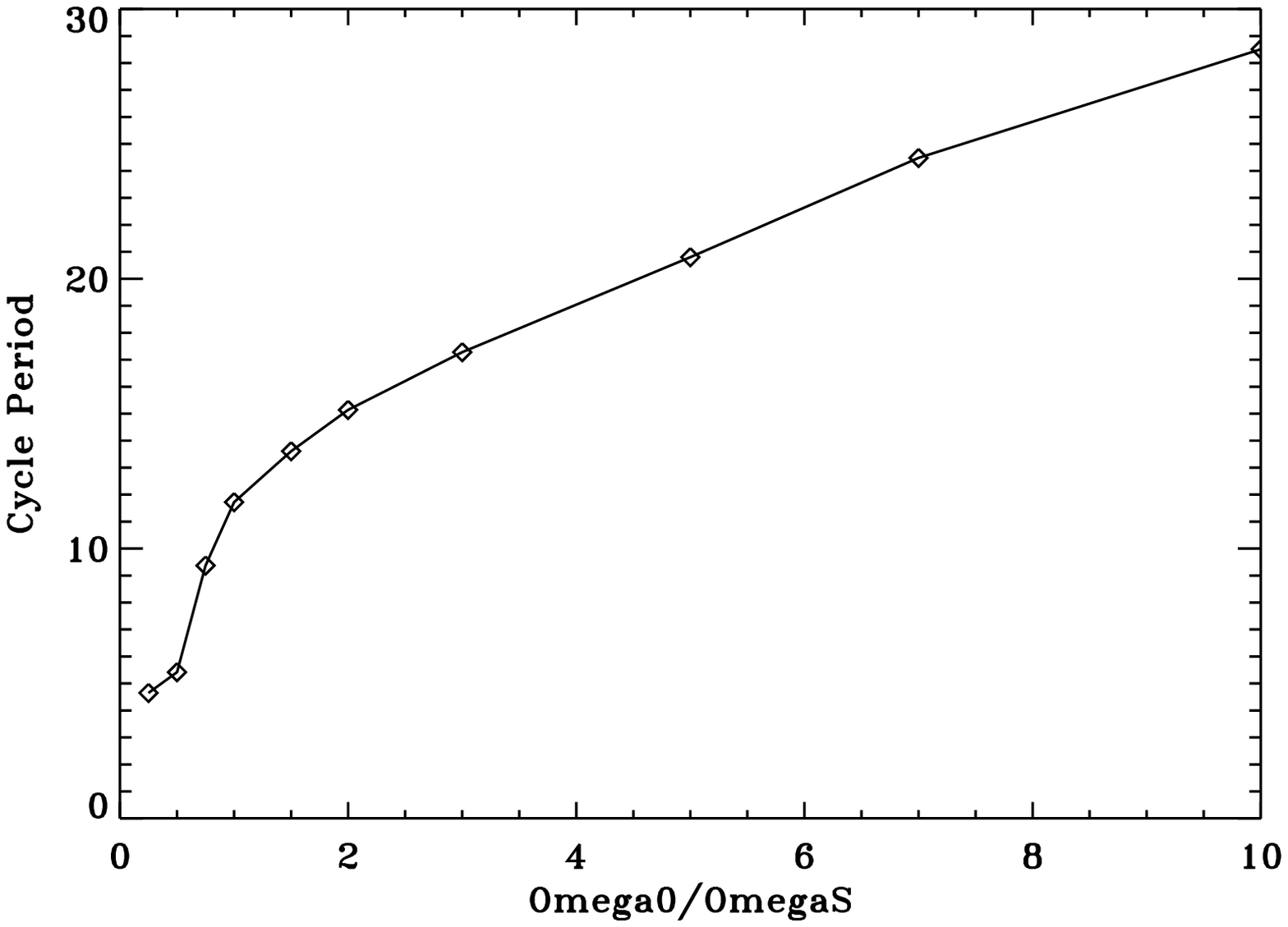}
\includegraphics[width=6.8cm]{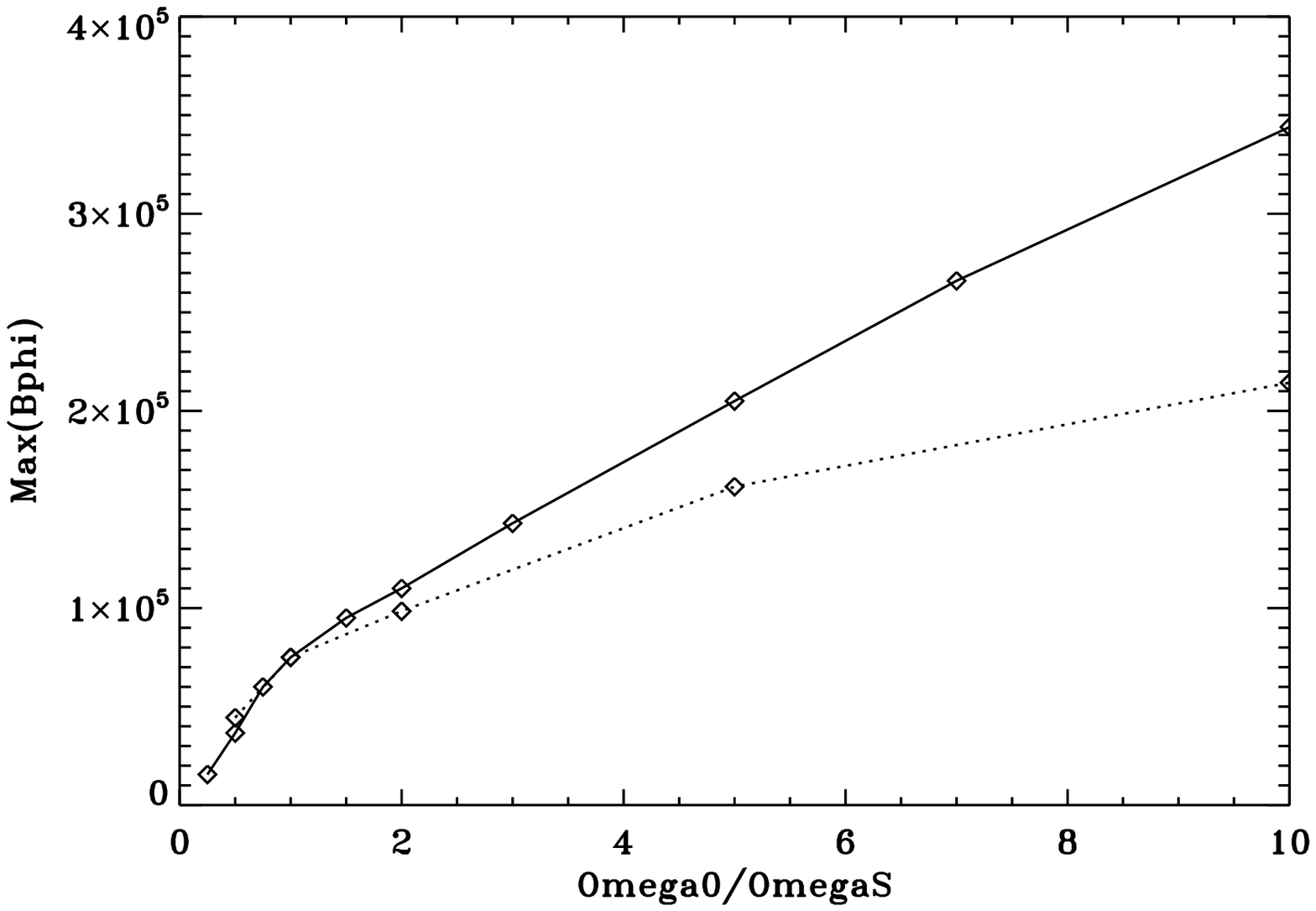}
\includegraphics[width=6.8cm]{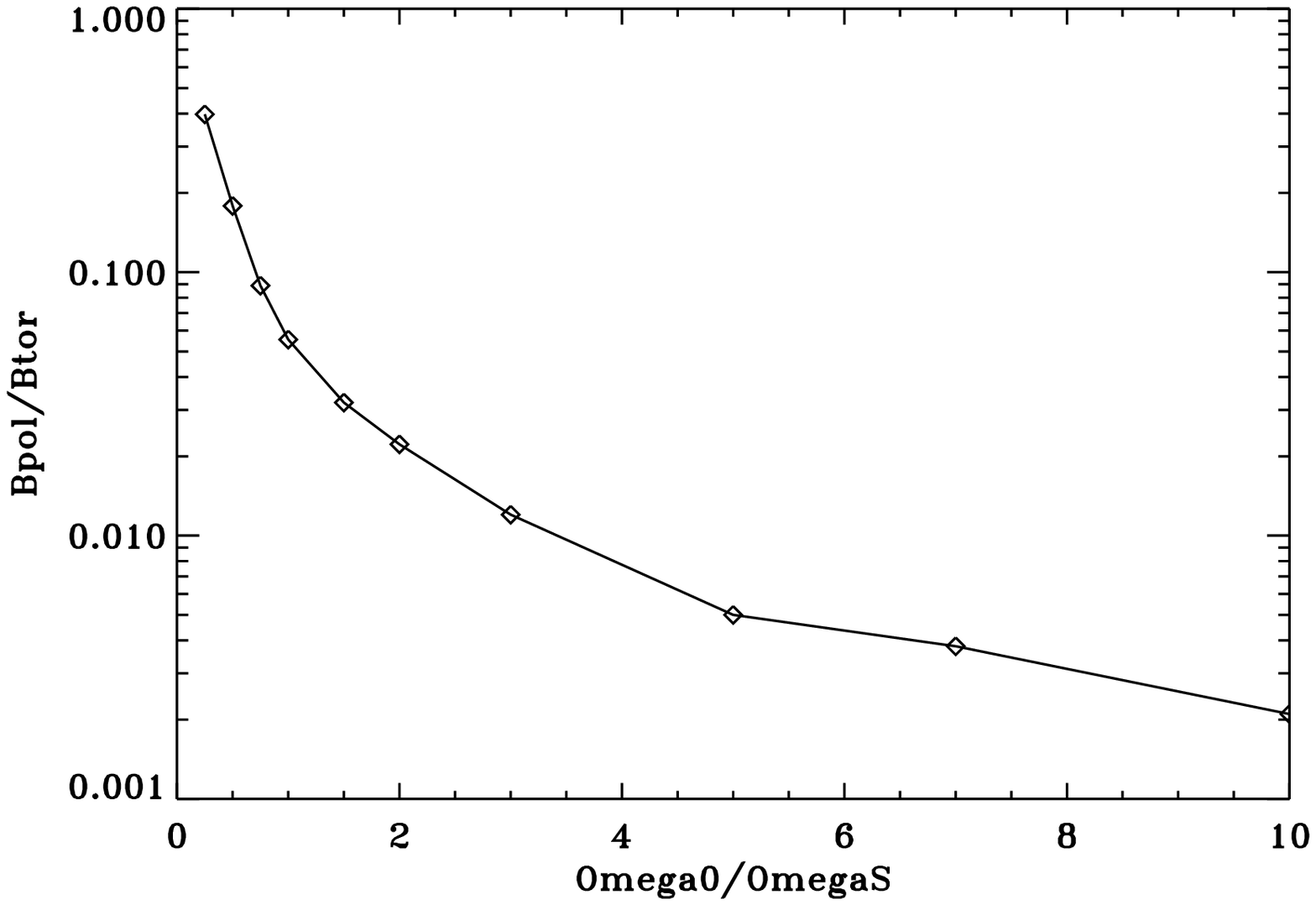}
\caption{Top panel: Variation of the magnetic cycle period wih respect to the amplitude of the MC and (2nd panel) to the rotation rate
for the cases G1 to G10. On the third panel we display the variation of the maximum 
of toroidal field at the base of the CZ for two series of models. The solid line corresponds to the models G1 to G10 using the scaled
down meridional flow as a function of the rotation rate whereas the dotted line corresponds to another set of models for which $v_0$ was kept constant.
On the bottom panel, we show the ratio Bpol/Btor wih respect to the rotation rate for cases G1 to G10.} 
\label{fig_courbes}
\end{figure}

It is clear that the evolution of the field topology as a function of the rotation rate is
of great importance. We are thus summarizing in Table \ref{table_courbes} 
and on the curves of figure \ref{fig_courbes} how the field intensity and topology is changing with $\Omega_0$ and the meridional circulation amplitude. 
We show for instance the variation of the magnetic cycle period with respect to the amplitude of the MC, with respect to 
the rotation rate and the variation of the maximum of toroidal field at the base of the CZ with respect to the rotation rate. 

\begin{table}[!h]
\begin{center}
\begin{tabular}{|c|c|c|c|c|} \hline 
\bf Case    & {\bf $\Omega_{0}/\Omega_{S}$} & {\bf Cycle period} & {\bf Max $B_\phi$} & Bpol/Btor \\ \hline
G0.25   &  0.25   &  4.65 & 1.56   & 0.398 \\ \hline
G0.5   &  0.5    &  5.42  & 3.66  & 0.179 \\ \hline
G0.75 & 0.75 & 9.37 & 6.00 & 0.0889\\ \hline
G1     &  1   &  11.7 & 7.50 & 0.0555\\ \hline
G1.5 & 1.5 & 13.6 & 9.50 & 0.0319 \\ \hline
G2       &  2      &  15.1 & 11.0 & 0.0222 \\ \hline
G3       &  3      &  17.3 & 14.3 & 0.0120 \\ \hline
G5       &  5      &  20.8 & 20.5 & 0.00500 \\ \hline
G7       &  7     &  24.5  & 26.6 & 0.00380 \\ \hline
G10      &  10    &  28.5  & 34.4  & 0.00210\\ \hline
\end{tabular}  
\caption{Starspot cycle period (in year), maximum of toroidal field (in $10^4 \, G$) at the base of the CZ and ratio between
poloidal and toroidal field  for 10 different cases with various rotation rates.}
\label{table_courbes}
\end{center}
\end{table}

More precisely on the first (top) panel of Figure \ref{fig_courbes}, we recover the almost proportional relationship between the magnetic cycle period and the inverse of 
the meridional flow amplitude. This again shows the major impact of the meridional circulation in these Babcock-Leighton dynamo models on the cycle period.
As stated in Brown et al (2008), the meridional circulation kinetic energy in 3-D hydrodynamical (HD) models of rapidly rotating stars seems to scale as approximately $\Omega_{0}^{-0.9}$,
leading the averaged meridional flow amplitude to scale as $\Omega_{0}^{-0.45}$. Indeed, on the second panel of Figure \ref{fig_courbes} where we show the cycle period plotted against the rotation rate of the star, 
we naturally recover a scaling in $\Omega_{0}^{0.45}$. A star rotating at $8 \Omega_{S}$ for example would have a cycle period of about 26 years approximately $4^{0.45}=1.87$ times 
the period of 15.1 years of case G2 rotating at twice the solar rate. 

More interestingly, if we now focus on the third and fourth (bottom) panels of Figure \ref{fig_courbes}, we note that the amplitude of the toroidal field possesses an almost linear 
dependency on the rotation rate $\Omega_0$. This property is due mainly  to two effects: a) a larger  $\omega$-effect (via its dependence on the Reynolds number $C_{\Omega}$) 
in more rapidly rotating stars and b) the fact that the reduction of the meridional flow amplitude  allows the poloidal field to be stretched and amplified over a longer period of time
(see Nandy (2004) and Yeates et al. (2008) who are finding similar effects). 
This is evident by comparing on the third panel of Figure \ref{fig_courbes} the solid and dashed curves. The difference between the two curves is the amplitude of
the meridional flow as a function of rotation rate that we kept constant for the cases shown using dashed line (these models
were used to determine the least square fit of equation (12)). We clearly see that both series of models lead to larger toroidal field as  we increase the rotation rate
but for the series with the scaled down meridional flow, the strengthening of $B_{\phi}$ is even higher. As a result, in our set of 2-D BL dynamo models the poloidal to toroidal
field ratio is found to increase almost linearly with the rotation period (inversely proportional to the rotation rate as seen on the last panel of figure \ref{fig_courbes}). 
This is in reasonable agreement with the observations of rapidly rotating solar-like stars by Petit et al. (2008). However the scaling of Bpol/Btor found in recent 3-D MHD 
simulations computed by Brown et al. (2009) is not linear. Moreover, the observations indicate that the energy stored in the large-scale toroidal
component of the magnetic field tends to dominate for rotation periods lower than 12 days (corresponding approximately to a rotation rate of twice the solar one). In our cases, the toroidal
field is always dominant, even for the case with $0.25\Omega_{S}$ where the poloidal field hardly reaches $40 \%$ of the toroidal field. 

This discrepancy could be due to the fact that the values
quoted in Petit et al. (2008) are related to observationally detectable fields, i.e. to surface fields; whereas the values we quote in this work are related to the magnetic fields in the 
whole convection zone. If the strong toroidal fields stored at the base of the convection zone are not able to reach the solar surface (as it is the case under certain conditions, 
see Jouve \& Brun 2009 for more details), the surface measurements could underestimate the amount of magnetic energy stored in the toroidal fields in the whole stellar convective envelope.

\section{Varying the model parameters}

From the above set of models it is clear that just increasing the rotation rate of the model and/or implementing a scaled meridional
circulation amplitude is not sufficient to reproduce the fast (close to linear) reduction of the stellar cycle period with the rotation rate. We thus wish in this section to modify all the other parameters
in order to find the best set of models to reproduce the observations.

The first easiest modification is to change the value of the diffusivity or of the constant $s_0$ used in the model. From the least square fit shown in equation (\ref{pcyc_theo}) 
the weak dependency of $P_{cyc}$ on $\eta$ or $s_0$ indicates that it is unlikely that such a modification will be able to reproduce the observations. We thus need to look for more sophisticated
solutions. In order to make the discussion easy to follow and to reduce the number of models considered we will now limit our study to cases rotating at five times the solar rate. Of course our results
are easily transposable to other rotation rates.

\subsection{Modifying the differential rotation profile}

\begin{figure}
  \centering
\includegraphics[width=8cm]{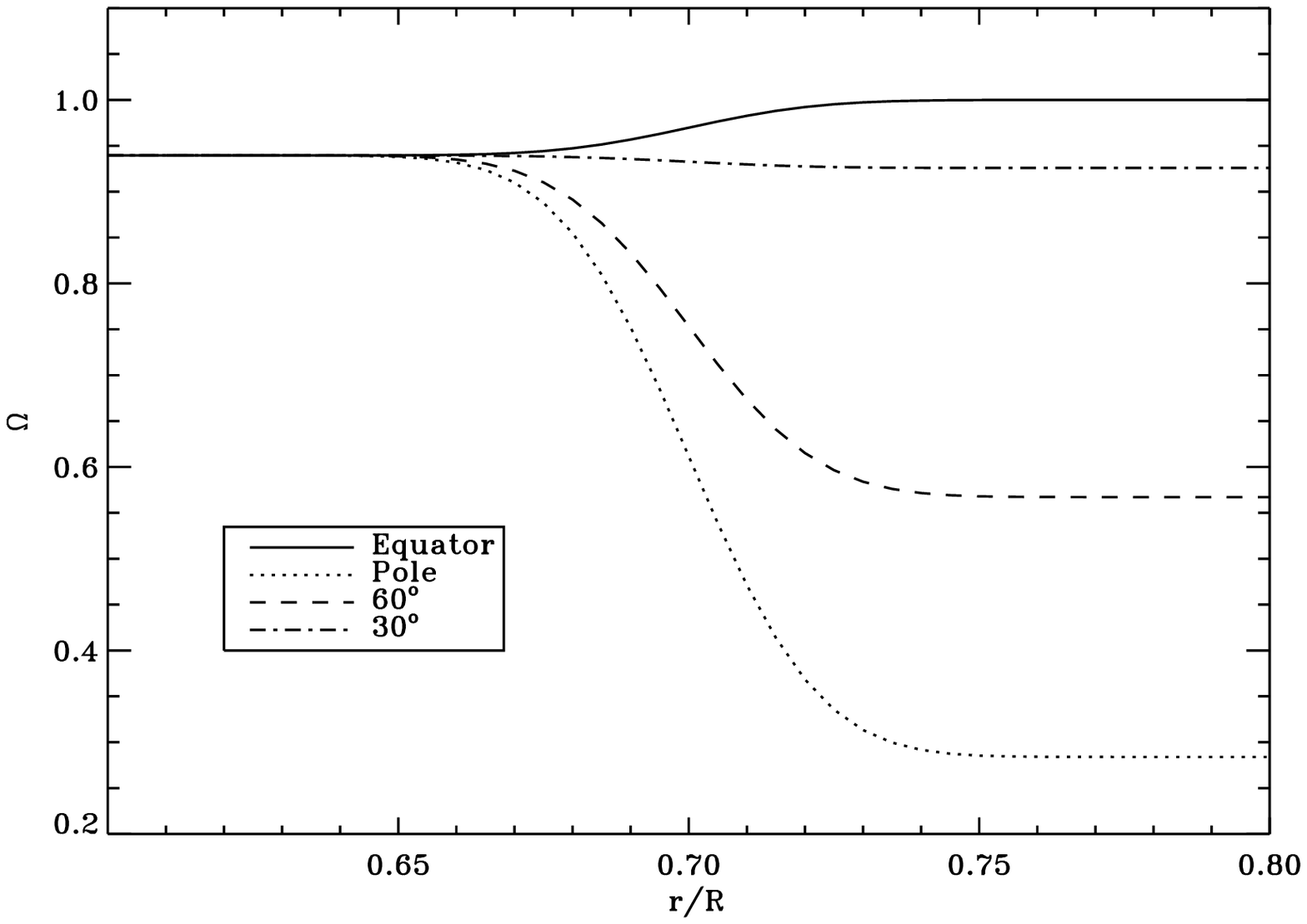}
\includegraphics[width=9cm]{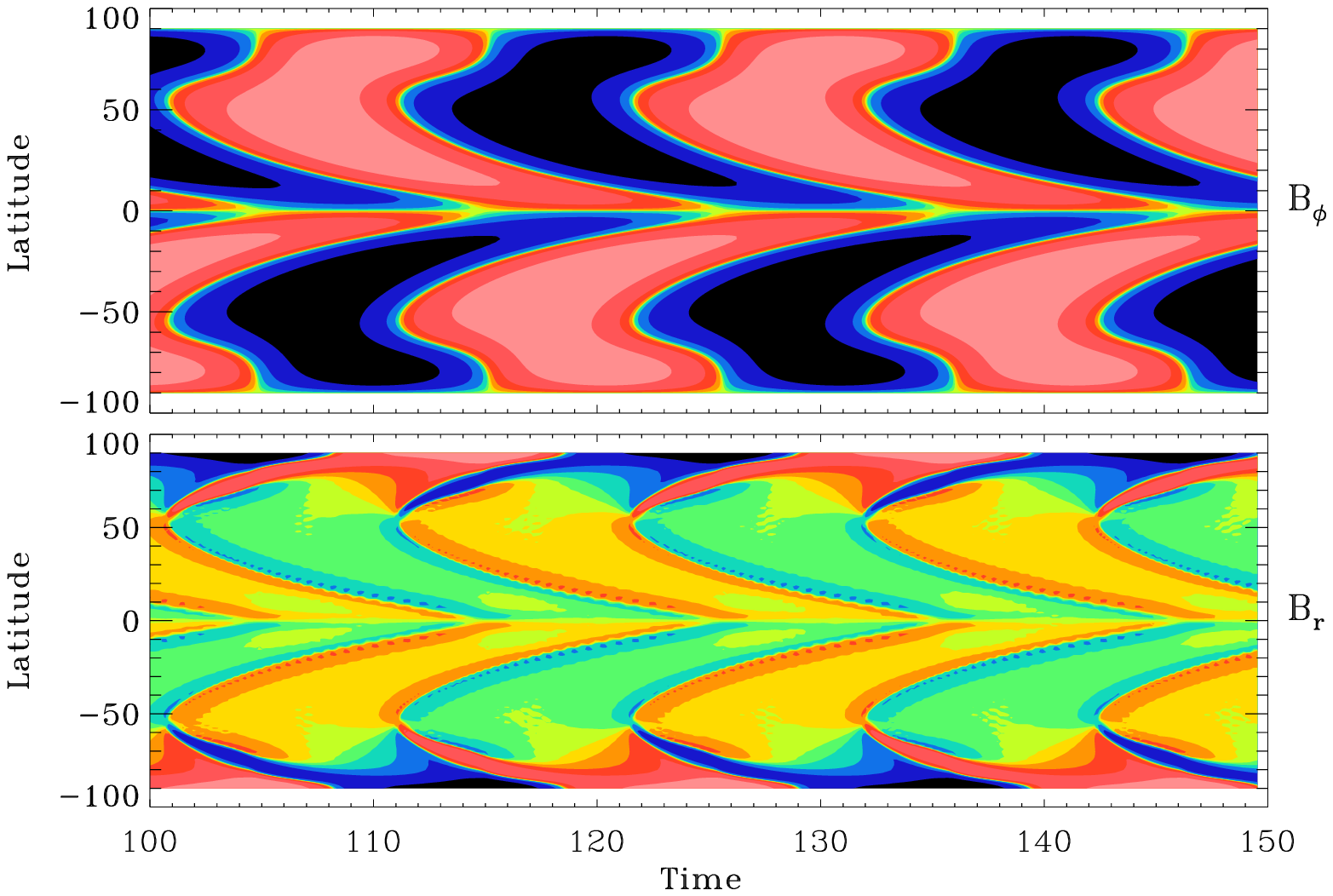}
\caption{Top panel: New differential rotation used in the model G5om zoomed on the portion between $r=0.6$ and $r=0.8$. For that case $\Delta \Omega$ has been increased by about a factor of 3 following Donahue et al. (1996) scaling. 
Bottom panel: Butterfly diagrams for that same case G5om with increased $\Delta \Omega$. The format is the same as Figure \ref{fig_butterfly}.} 
\label{fig_butterfly_domega}
\end{figure}

As discussed in the introduction it is clear that in low mass stars the differential rotation is expected to change with the rotation rate. Given the key role played
by the $\omega$-effect in mean field dynamo, we propose to compute a new model at 5 times the solar rate with a modified differential rotation, namely case G5om. But what scaling 
of $\Delta \Omega$ with $\Omega_0$ should we use?
In their 3-D hydrodynamic models, Brown et al. (2008) find a scaling of $\Delta \Omega \propto \Omega_0^{0.3}$. However due to the nonlinear feed back of the Lorentz force on the mean longitudinal flow, this scaling is not recovered in the MHD models computed in Brown et al. (2009). We thus choose here not to use the theoretical scaling for $\Delta \Omega$ of Brown et al. (2008) since it is somewhat uncertain and depends on whether or not dynamo action is realized in the simulations.
The observational situation is not clearer with either a weak scaling close to zero (i.e. the differential rotation does not vary with the rotation rate; Barnes et al. 2005) or strong scaling as in Donahue et al (1996). Given the current debate going on among the observers on the exact variation of $\Delta \Omega$ with $\Omega_0$ we have some freedom in the choice of our variation. We choose here to follow the scaling proposed by Donahue et al. (1996), i.e. $\Delta \Omega \propto \Omega_0^{0.7}$ since it is the most extreme variation found among the observers and as a
consequence it will maximize the effect of varying the differential rotation profile. The new differential rotation profile is shown on the top panel of Figure \ref{fig_butterfly_domega} and is directly comparable with the one shown on the top panel of Figure \ref{domega}.
We also display on the bottom panel of Figure \ref{fig_butterfly_domega} the butterfly diagram for this new model G5om rotating at 5 times the solar rate with increased $\Delta \Omega$.

We see that the cycle period is almost not affected by the increase of the latitudinal shear when comparing the butterfly diagram of Figure  \ref{fig_butterfly_domega} with the third panel of Figure \ref{fig_butterfly}. 
This confirms that the $\omega$-effect plays little role in setting the cycle period in BL dynamo models.
On the contrary the ratio between Bpol and Btor is changed. Indeed, for this case, we go from a Bpol to Btor ratio of 0.05 in the standard case G5 to a ratio of 0.019 in G5om. 
We thus proceeded in this case to a significant increase of the toroidal component of the magnetic field, keeping the value for the poloidal field approximately the same.
 This is due to the fact that this new class of models is generating stronger Btor as can be expected with a larger shear. A precise observation of this ratio may thus be of utility to put constraints on stellar internal differential rotation. 

With a larger latitudinal shear it is expected that the tachocline at the base of the stellar convection zone will also be modified. The tachocline is
the transition region between the differentially rotating convective envelope of low mass star and their supposedly uniformly rotating radiative interior. How
the tachocline thickness is varying with the rotation rate and the differential rotation of the star is not easy to determine. Based on the
tachocline model of Spiegel and Zahn (1992) and the observations of Donahue et al. (1996) for the dependency of $\Delta\Omega$ with $\Omega_0$, Brun et al. (1999) have shown that the tachocline thickness for a younger more rapidly 
rotating Sun could be thicker. They find a scaling for $d \propto \Omega_0^{(1.7\pm0.1)/4}$ when taking into account all the dependencies. We have thus computed a new model G5omt,
with both a larger tachocline (about twice as large in this case than in case G5om, following the scaling law quoted above) and stronger differential rotation (identical to case G5om shown on Fig. \ref{fig_butterfly_domega}, with a $\Delta\Omega$ about three times the solar one). The result in term of the cycle period is again difficult to distinguish from the standard G5 model, that had only $C_{\Omega}$ and the meridional circulation amplitude changed. This confirms that the
tachocline plays almost no role in Babcock-Leighton dynamo models in setting the cycle period.
On the contrary the ratio Bpol/Btor is modified being intermediate between the cases G5 and G5om. The value of the ratio is now of about 0.02, very close to the case where the differential rotation was increased, but with a slight reduction of the toroidal component, due to the softer transition (i.e. thicker tachocline). In summary neither models G5om nor G5omt possess a
cycle period in agreement with the observations and we thus need to modify another physical ingredient.

In the standard Babcock-Leighton model we are thus left with two possibilities if we want to
reconcile the models and the observations: modifying the profile of meridional circulation or introducing the effects of a faster rotation rate on the surface source term. Indeed, for example it has been shown by D'Silva \& Choudhuri (1993) that the Coriolis force could strongly act on the tilt angle of emerging active regions, thus modifying the source term for poloidal field. This feature will have to be checked in future works, with the help of 3-D MHD models such as the ones of Jouve \& Brun (2009). For this work, since the dependence of the cycle period on the amplitude of the meridional flow is very strong, we decide to focus on the first solution: modifying the MC profile.

\subsection{Varying the number of meridional circulation cells}

The strong relation between the meridional circulation speed and the cycle activity is an indication that shortening the advection path may result
in shorter cycle period. Dikpati et al. (2004), Bonanno et al. (2005) and Jouve \& Brun (2007a) have studied in detail the influence of various meridional circulation profiles
on the butterfly diagram and other large scale magnetic properties of the Sun. 

\begin{figure}[!h]
  \centering
\includegraphics[width=9cm]{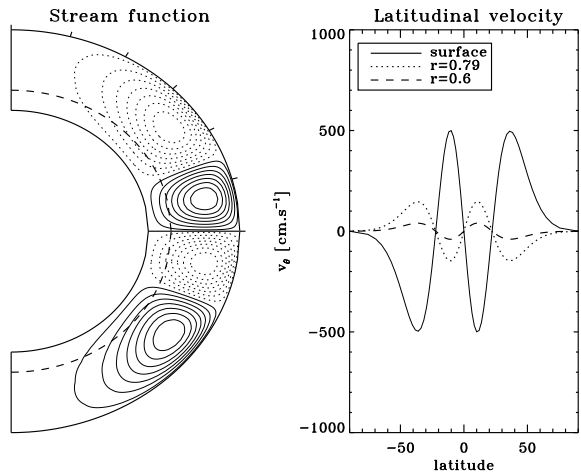}
\caption{Multicellular stream function for the case rotation at 5 times the solar rate G5mc20. Two cells with latitude have been used, with
the first counterclockwise cell near the equator closing at 20 deg of latitude. On the left panel we show the streamfunction and on the right panel the latitudinal velocity at 3 depths.} 
\label{fig_new_mc}
\end{figure}

In particular Jouve \& Brun (2007a) have shown that adding cells in radius slows down the cycle whereas
as the two other groups also showed, adding cells in latitude accelerates it. If one seeks to shorten the cycle length as it is the case in this study, only adding cells in latitude will have
the expected impact. We have thus computed a new set of simulations, rotating at five times the solar rate in which we have only modified the meridional circulation in latitude (leaving all the other parameters the same as in case G5) by considering two cells in latitude per hemisphere with various extension instead of just one cell.

Following the 3-D results of Brown et al. (2008) and particularly figure 11 of their article, we choose to introduce a meridional circulation with a counterclockwise cell at lower
latitudes and a clockwise cell at higher latitudes in the Northern hemisphere, this profile being antisymmetric with respect to the equator. Since uncertainties are significant concerning
the extension of the low-latitude meridional cell, we have computed several models at 5 times the solar rate including a low-latitude cell extending to $45{\degr}$, $40{\degr}$, $30{\degr}$ and $20{\degr}$, named respectively G5mc45, G5mc40, G5mc30 and G5mc20. For the sake of clarity we have decided to illustrate and discuss in length in this section only one case, namely case G5mc20, the model with the less extended lower meridional circulation cell. The meridional circulation profile for that case is shown on Figure \ref{fig_new_mc} for illustration. In all models we have set the latitudinal velocity such that it has the same maximal amplitude in the two counter-cells which correspond to that in the case with one single cell (i.e. case G5) keeping all the other parameters the same. 

\begin{figure}[!t]
  \centering
\includegraphics[width=9cm]{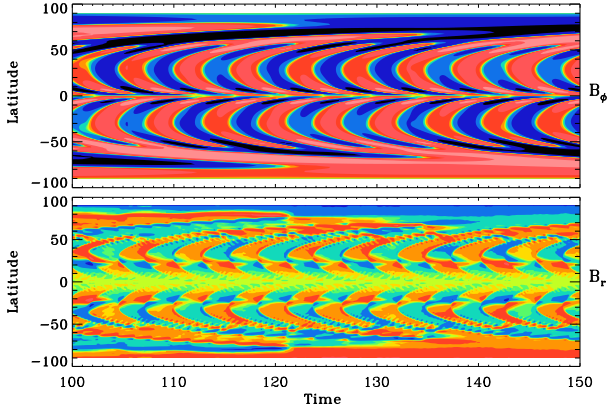}
\caption{Butterfly diagram of the case G5mc20, rotating at 5 times the solar rotation rate and possessing a multicellular meridional flow as shown on Figure \ref{fig_new_mc}.} 
\label{fig_new_mc_pap}
\end{figure}

\begin{figure}[!t]
  \centering
\includegraphics[width=9.2cm]{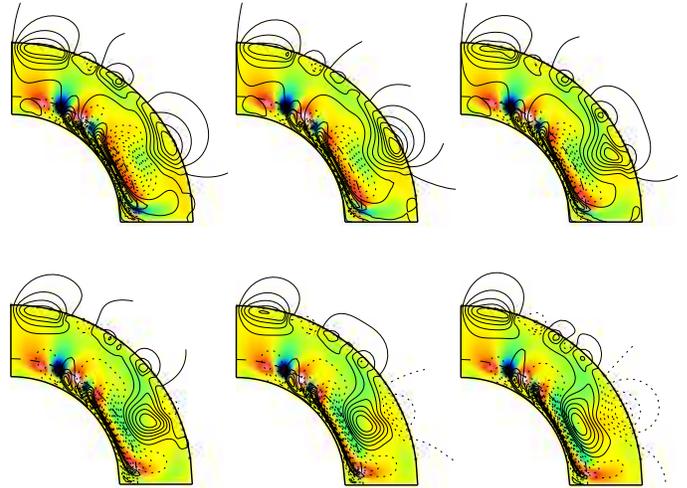}
\caption{Same as Figure \ref{fig_solution} but for the case G5mc20 rotating at 5 times the rotation rate and possessing a multicellular meridional circulation as shown on Figure \ref{fig_new_mc}.
The evolution is shown over a temporal interval of 5.2 years corresponding to half a magnetic cycle or one starspot cycle.} 
\label{fig_new_mc_evol}
\end{figure}

When comparing the various multi-cellular stellar dynamo simulations, we find that the scaling of the cycle period with the extension of the low-latitude cell is almost linear, the starspot cycle period being respectively of 9.9, 9.2, 7.4 and 5.2 years for those different models. This is mainly due to the fact that the advection path is reduced by the presence of counter-cells and will shorten the distance each component of the magnetic field has to travel before reaching the region where it will be turned into the other component. According to the previous studies and scaling laws of  Noyes et al. (1984), Saar \& Brandenburg (1999), Charbonneau \& Saar (2001) and Saar (2002), the cycle period for a star rotating at 5 times the rotation rate should be around 3 years. We are getting very close to this value with a low-latitude cell extending to $20{\degr}$ in latitude as in case G5mc20. We now focus on butterfly diagram and the evolution of the field in the meridional plane of this model, as shown on Figures \ref{fig_new_mc_pap} and  \ref{fig_new_mc_evol}.

On the butterfly diagram shown on Figure \ref{fig_new_mc_pap}, we clearly see that we have significantly reduced the length of the activity (starspot) cycle: the period is now 5.2 yr instead of 22 yr. The cyclic activity tends now to be more concentrated at low latitudes. The temporal evolution of the field seems to be less steady if we consider the variations of intensity which appear from one cycle to the next, visible on the butterfly diagram through changes in color between 
the different activity cycles. The radial field appears to be even more busy, exhibiting a periodic but yet complex pattern at low latitudes. We moreover note
that above $55\degr$, the cyclic activity is lost and we are left with some regions where the magnetic field evolves much more slowly than the
time scale of the main cycle. 

Looking at Figure \ref{fig_new_mc_evol} will help us to have more insight on how the field is organized in the convective envelope and how it is advected 
by the new meridional circulation. Again, we clearly see that the evolution of the field is much more complicated than in the standard case. We can
still follow the evolution of each component of the field on the various panels of the figure. 
Indeed we can distinguish the effect of the low-latitude cell which advects the newly generated poloidal field (counterclockwise on the first panel) downwards while the toroidal
field of the previous cycle (positive i.e. represented by reddish colours) is being advected at the base of the convection zone to reach the equator and replace the 
toroidal field of the previous cycle at the end of the starspot cycle (last panel). We note, in agreement with the butterfly diagram, that the advection at higher latitudes 
is much slower and as a consequence we do not clearly see any imprint of the activity cycle in the polar regions. This is due to the fact that the same amplitude of the flow 
is introduced for both cells and that the distance traveled by the field at low latitudes is much shorter, thus leading to a difference in the time-scale for the advection of
the magnetic field in the two cells. We also wish to reemphasize that we did not introduce multi cells in radius since Jouve \& Brun (2007a) showed that it results in a slowing down of the cycle period rather than a speeding up as required in this study
for fastly rotating stars. In summary adding cell in latitude clearly helps shortening the cycle for fastly rotating stars. For the stars rotating slower than the Sun it may be necessary to add cells in radius if one wants to increase the cycle. Indeed 3-D models
indicate that slower the star rotates faster its meridional circulation is and thus shorter its cyclic period will be as demonstrated by the butterfly diagram shown on the top panel of Figure \ref{fig_butterfly} for case G0.5.

\section{Summary and Conclusions}

The idea of this work was to better understand stellar magnetism, using both 2-D and 3-D calculations. To do so, we have introduced 
into Babcock-Leighton (B-L) flux transport dynamo models, results from 3-D global hydrodynamical simulations  by Brown et al. (2008). 
To be more specific we have used these simulations to determine the variation of the amplitude of the meridional circulation with rotation rate 
and complemented our study with observations of the variation of the angular velocity contrast $\Delta \Omega$ 
with rotation rate. Using these scaling laws  we have built a series of 2-D mean field stellar dynamo models in which
we have modified in turn the rotation rate, the amplitude of the meridional circulation and of the angular velocity gradient. 
This has allowed us to test the validity of this type of models on solar-like stars with faster rotations and intense magnetism.

Our study shows that an increase in the rotation rate leads to a larger $\omega$-effect which in turn modifies the ratio between 
poloidal and toroidal fields with a trend similar to that found observationally in a 
small sample of rapidly rotating suns (Petit et al. 2008). 
Other authors have also studied such trends using solar mean field B-L flux transport dynamo models (see for instance Charbonneau \& Saar 2001; Dikpati et al. 2001; Nandy \& Martens 2007).
They also find that the toroidal field grows in amplitude with the star's rotation rate.
The difference between our work and previous studies comes from the choice made in varying the meridional circulation amplitude with rotation rate. In 3-D global calculations an increase of the rotation rate 
leads to a decrease in the meridional flow amplitude whereas in previous studies the opposite was assumed since such knowledge was not yet available. 
Given that the cycle period in advection-dominated B-L dynamo models is almost inversely proportional to the intensity of the meridional circulation, 
we find that our models predict slower magnetic cycles for more rapidly rotating stars. This is not consistent with the observational 
data collected at Mount Wilson since the 1960's (e.g. Noyes et al. 1984, Baliunas et al. 1995).
It thus seems difficult to reconcile advection-dominated B-L models widely used in the solar dynamo community with stellar activity observations if we use scaling laws relating
rotation rates and meridional flow amplitudes given by consistent 3-D models. Of course the systematic observation of the meridional circulation realized in solar-like stars rotating at various rotation rate would
be of great use and could solve the apparent discrepancy if we indeed found that such flows are not so sensitive to the rotation rate. Unfortunately such observations are not yet available.
Hopefully with the Corot and Kepler satellites now in orbit this situation may change in the near future.

We have thus tried to investigate ways to improve the agreement between 2-D models and observations in refining our calculations. A least-square fit on the various parameters of the standard model
shows that we cannot rely on a modifications of $\Omega_0$, $s_0$ and $\eta$ to act on the cycle period, as the latter is almost completely determined by the amplitude of the meridional flow (see Equation 12). 
We thus computed more sophisticated models where the differential rotation and meridional circulation profiles were modified.

First, we computed a model at 5 times the solar rotation rate where the modified contrast
in the angular velocity was taken into account. Some uncertainties both from an observational or a theoretical standpoint still exist 
concerning the scaling between $\Delta\Omega$  and $\Omega_0$ but we find that the effect on the cycle
period is negligible, even when we consider the most extreme observation of the variation of the differential rotation contrast with rotation rate (Donahue et al. (1996)). 
Modifying the tachocline thickness as a function of rotation rate according to the scaling law of Brun et al. (1999)
does not help to improve the relationship between the rotation rate of the star and the cycle period either. We show that the main impact of a modification
of the profile of differential rotation is to decrease the percentage of the poloidal component in the total magnetic field again because of the increased efficiency of the
source for toroidal field: the $\omega$-effect. The modification of the source term for poloidal field (decay of active regions in Babcock-Leighton models and the kinetic
helicity of the flow in $\alpha$-$\omega$ models) due to an increase of the rotation rate still needs to be addressed to study the evolution of the poloidal to toroidal field ratio 
in those refined cases.

Given the strong dependency of the magnetic cycle period on the meridional circulation and the fact that 3-D models exhibit multicellular patterns of the flow in latitude, it is
reasonable to think that modifying the profile of the MC according to 3-D models may help to get closer to the observational data. Indeed, we show that adding a counter cell in latitude
which extends until about $20\degr$ in both hemispheres reduces drastically the cycle period. We find a quasi-linear relationship between the extension of the counter cell and the cycle
period, which is very promising to reconcile Babcock-Leighton models with observations of stellar activity. 
Another solution could be to change the way poloidal magnetic fields are transported from the surface down to the tachocline. 
In Yeates et al. (2008) the transport is dominated  by diffusion to link the distant magnetic field generation regions. In their case the cycle period seems less sensitive to the meridional flow 
amplitude and profile assumed and depends more on the diffusion profile. Guerrero \& Gouveia Dal Pino (2008) have introduced in Babcock-Leighton dynamo equations 
an extra term representing the pumping of magnetic fields due to turbulent convection (inspired by the work of K\"{a}pyl\"{a} et al. 2006). This new process leads to vertical and 
horizontal transport of the fields. Their model seems to show a significant decrease
of the cycle period dependency on the meridional flow amplitude (from $v_0^{-0.9}$ to $v_0^{-0.12}$, although only a shallow meridional flow is considered,
 which may influence the scaling law) and that the turbulent pumping is now responsible for regulating the cyclic activity. 
Reintroducing the meridional flow amplitude scalings deduced from 3-D calculations in this type of models would thus have much less impact on the cycle period. The 
changes in the pumping coefficients with respect to the rotation rate would have to be investigated to see if some of the observations of stellar magnetic activity could be recovered.

 In conclusion, advection-dominated Babcock-Leighton flux-transport models may offer 
 insights into the dynamo action occurring within the convection 
 zones of solar-type stars.  However, it is necessary to make 
 serious modifications to these models if we wish to reconcile 
 their predictions with observations of real stellar magnetism.

\begin{acknowledgements}
We acknowledge funding from the European Community via the ERC-StG grant STARS2 207430 (www.stars2.eu).
We are thankful to T. Emonet \& P. Charbonneau for the original version of the STELEM code, to M. Browning for fruitful discussions  
on stellar magnetism and rotation and to the organizers of the dynamo program at KITP in Santa Barbara in 2008 where this work was initiated.
We are also thankful to the anonymous referee whose comments helped improve the quality of the paper.
B.P Brown wishes to acknowledge fundings through the NASA GSRP program by award number NNG05GN08H.
\end{acknowledgements}

\end{document}